\documentclass[%
reprint,
superscriptaddress,
amsmath,amssymb,
aps,
floatfix,
]{revtex4-2}

\usepackage[utf8]{inputenc}
\usepackage{xcolor}
\usepackage{graphicx}
\usepackage{subfigure}
\usepackage{hyperref}
\hypersetup{colorlinks,allcolors=blue}

\begin{document}

\title{Diffuse Maxwellian illumination for safe wide-field retinal Doppler holography}

\author{Zofia Bratasz}
\author{Olivier Martinache}
\author{Yohan Blazy}
\author{Ang\`ele Denis}
\author{Coline Auffret}
\author{Jean-Pierre Huignard}
\affiliation{
Centre National de la Recherche Scientifique (CNRS) UMR 7587, Institut Langevin. Paris Sciences et Lettres (PSL) University, Sorbonne Universit\'e (SU). \'Ecole Sup\'erieure de Physique et de Chimie Industrielles (ESPCI) Paris - 1 rue Jussieu. 75005 Paris. France
}

\author{Ethan Rossi}
\author{Jay Chhablani}
\author{Jos\'e-Alain Sahel}
\affiliation{
University of Pittsburgh, 203 Lothrop Street, Suite 800, Pittsburgh, PA, 15213, USA.
}

\author{Sophie Bonnin}
\author{Rabih Hage}
\author{Patricia Koskas}
\author{Damien Gatinel}
\author{Catherine Vignal}
\author{Am\'elie Yavchitz}
\author{Ramin Tadayoni}
\author{Vivien Vasseur}
\affiliation{
Rothschild Ophthalmologic Foundation, Clinical studies department, 75019, Paris, France
}

\author{Claire Ducloux}
\author{Manon Ortoli}
\author{Marvin Tordjman}
\author{Sarah Tick}
\author{Sarah Mrejen}
\author{Michel Paques}
\affiliation{
Quinze-Vingts National Eye Hospital, DHU Sight Restore, Sorbonne Universit\'e, INSERM-DGOS CIC 1423, CNRS, 28 rue de Charenton, Paris, 75012, France.
}

\author{Michael Atlan}
\affiliation{
Centre National de la Recherche Scientifique (CNRS) UMR 7587, Institut Langevin. Paris Sciences et Lettres (PSL) University, Sorbonne Universit\'e (SU). \'Ecole Sup\'erieure de Physique et de Chimie Industrielles (ESPCI) Paris - 1 rue Jussieu. 75005 Paris. France
}

\date{\today}

\begin{abstract}
We report a diffuse Maxwellian illumination scheme for wide-field retinal laser Doppler holography. Inserting an engineered diffuser in the illumination arm transforms a spatially concentrated near-infrared laser focus into an angularly diversified illumination pattern, thereby reducing local irradiance near the anterior segment while preserving coherent interferometric detection. This configuration allows the eyepiece to be positioned closer to the cornea, increasing the digitally reconstructed retinal field of view without producing a localized corneal hot spot. We compare three illumination geometries: focused non-diffuse illumination, diffuse illumination at the same cornea--eyepiece distance, and diffuse Maxwellian illumination. Diffuse Maxwellian illumination expands the retinal field of view while preserving Doppler contrast in broad and high-frequency fluctuation bands. Light-hazard assessment is limited to the current ophthalmic standards ISO 15004-2:2024 and ANSI Z80.36-2021. Based on measured beam profiles, the recommended operating power at 852 nm is set by the most restrictive relevant exposure condition among the assessed anterior-segment, iris, and retinal limits. These results support diffuse illumination as a practical route toward safer, non-mydriatic, wide-field Doppler holography of the human retina.
\end{abstract}

\maketitle

\section{Introduction}

Optical holography methods increasingly leverage laser radiation for coherent, phase-resolved computational imaging in ophthalmology. In retinal laser Doppler holography, camera recordings of interferograms formed by the light backscattered by the retina and a separate reference beam are used to compute local optical Doppler contrasts \cite{Puyo2018}. This imaging scheme enables Doppler imaging of the posterior segment and can also provide anterior-segment information from the same raw interferogram data set \cite{Puyo2021}.

A practical limitation of wide-field retinal Doppler holography is the local irradiance created when near-infrared laser light is focused near the anterior segment. In earlier implementations, the eye was typically illuminated with approximately 2 mW or less of continuous near-infrared laser exposure, with the illumination focus positioned in front of the cornea at a distance chosen according to the desired field of view \cite{Puyo2021}. When the focus is positioned in the eye front focal plane, the iris tends to act as an image field diaphragm rather than an aperture stop, which limits wide-angle retinal imaging. In practice, full posterior-pole imaging previously required rendering and stitching several sequential acquisitions \cite{Puyo2020}.

Maxwellian-view illumination is attractive for wide-field retinal imaging because a converging beam focused near the eye's nodal point can illuminate a large retinal area \cite{Sliney2005}. However, in a coherent laser system, this configuration can create high local irradiance near the anterior segment. We therefore introduce an engineered diffuser in the illumination path to redistribute the optical power spatially and angularly before the eye. This reduces the localized hot spot while preserving enough coherence for heterodyne holographic detection.

In this work, we describe a diffuse Maxwellian illumination configuration for retinal Doppler holography, compare Doppler images obtained with and without the diffuser, and assess optical exposure using only the current ophthalmic light-hazard standards ISO 15004-2:2024 and ANSI Z80.36-2021. Older standards, including ISO 15004-2:2007 and ANSI Z136.1-2014, are not used for the main light-hazard analysis.

\section{Experimental setup}

\begin{figure}[htbp]
\centering
\includegraphics[width=\linewidth]{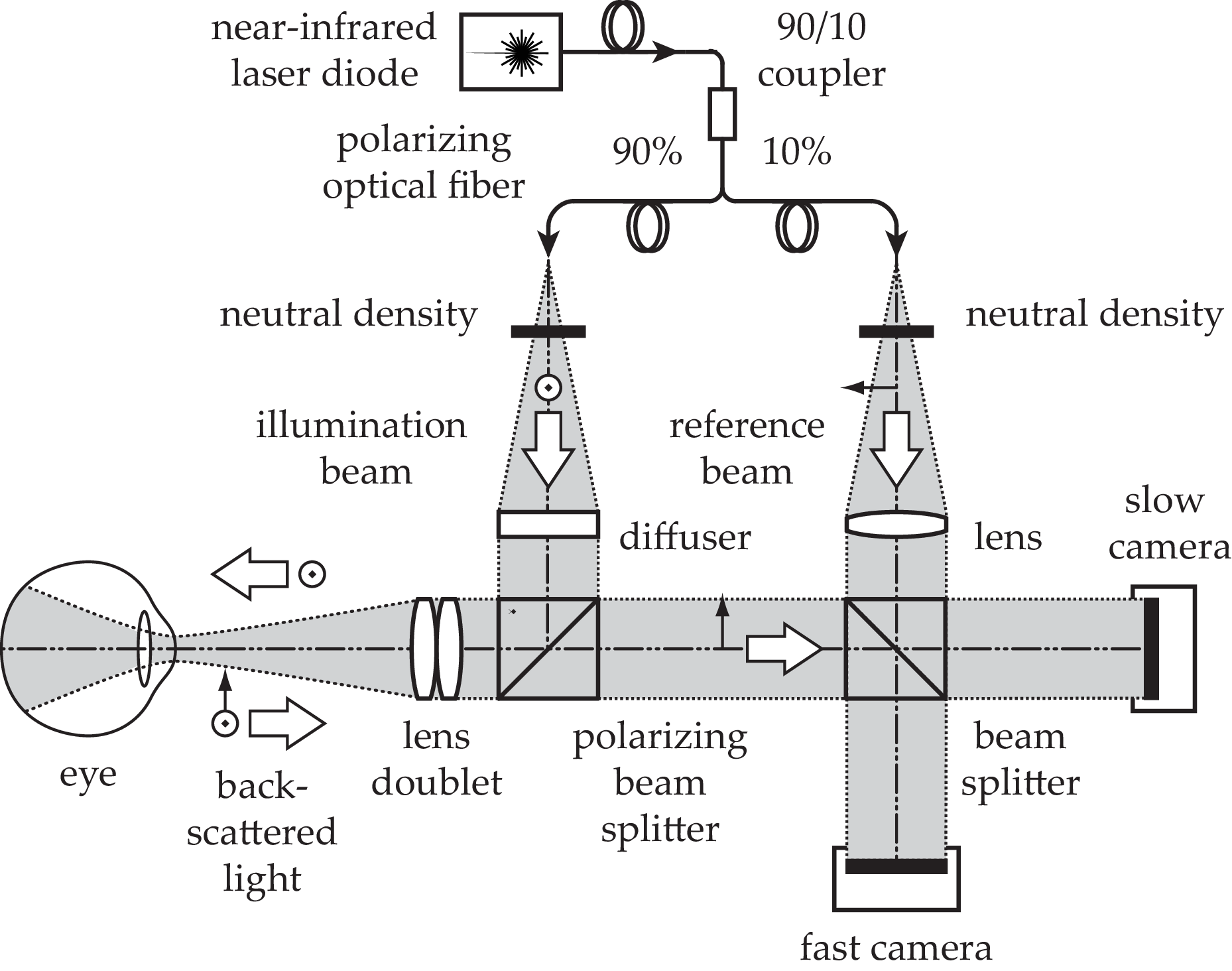}
\caption{Sketch of the optical configuration. An inline Mach--Zehnder near-infrared laser interferometer mixes light backscattered by the eye fundus of a volunteer with a separate reference beam. Two cameras record output optical interference patterns. The main difference with previously reported arrangements \cite{Puyo2018} is the presence of an engineered optical diffuser that scatters the illumination beam. The diffuser-to-eyepiece center distance is $\sim$100 mm.}
\label{fig_Setup}
\end{figure}

\begin{figure}[htbp]
\centering
\subfigure[]{\label{eyeConfig1}\includegraphics[width=0.68\linewidth]{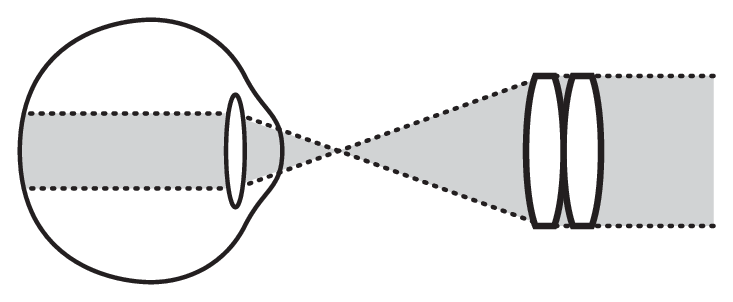}}
\subfigure[]{\label{eyeConfig1_Doppler}\includegraphics[width=0.28\linewidth,pagebox=mediabox]{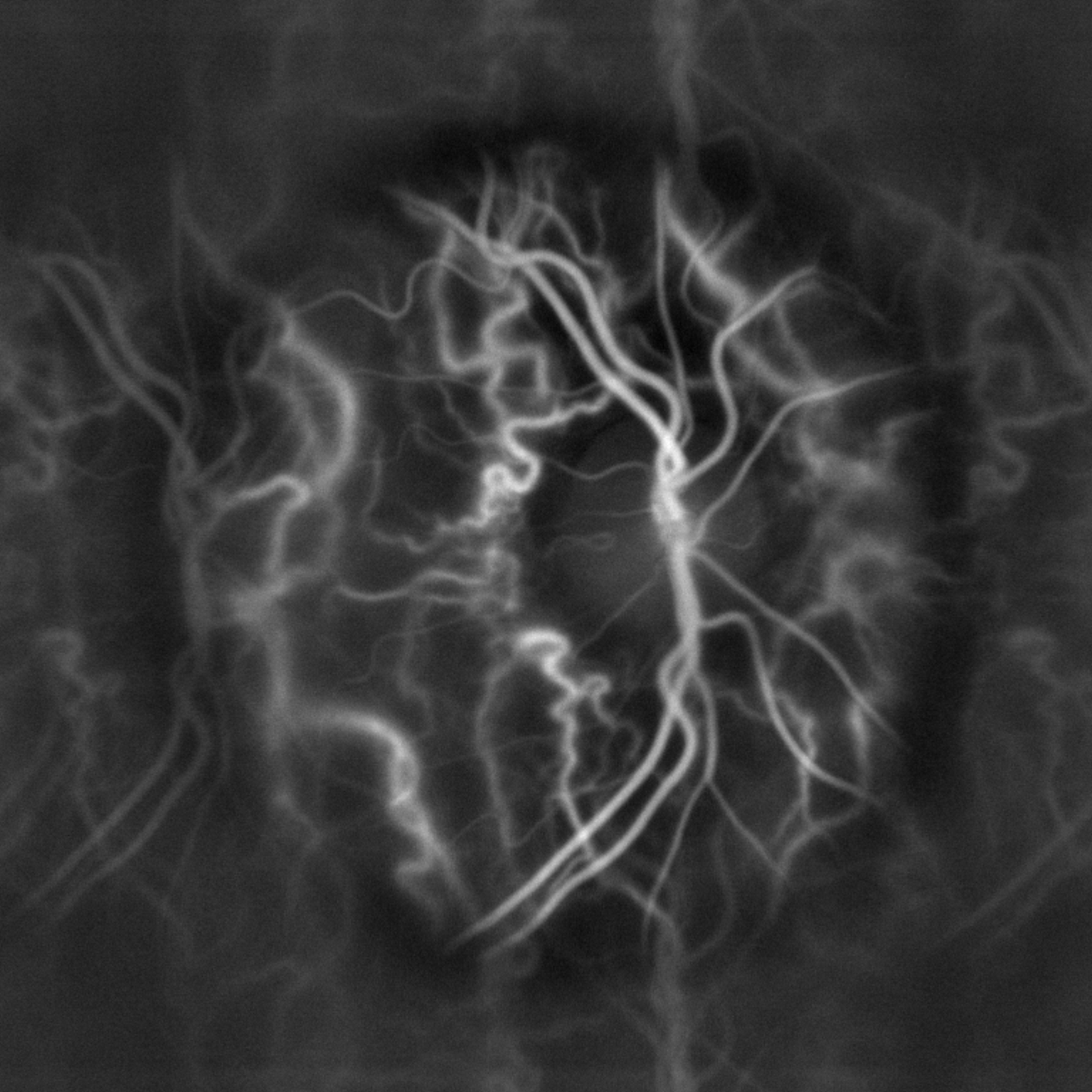}} \\
\subfigure[]{\label{eyeConfig2}\includegraphics[width=0.68\linewidth]{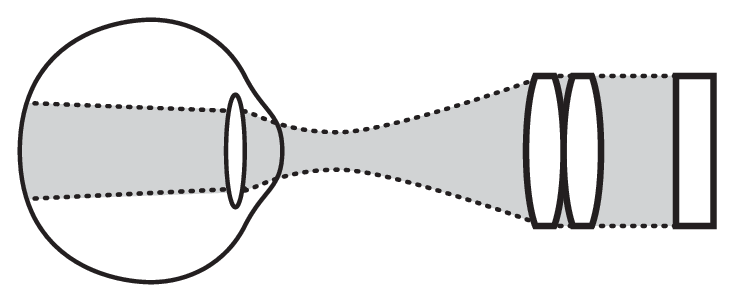}}
\subfigure[]{\label{eyeConfig2_Doppler}\includegraphics[width=0.28\linewidth,pagebox=mediabox]{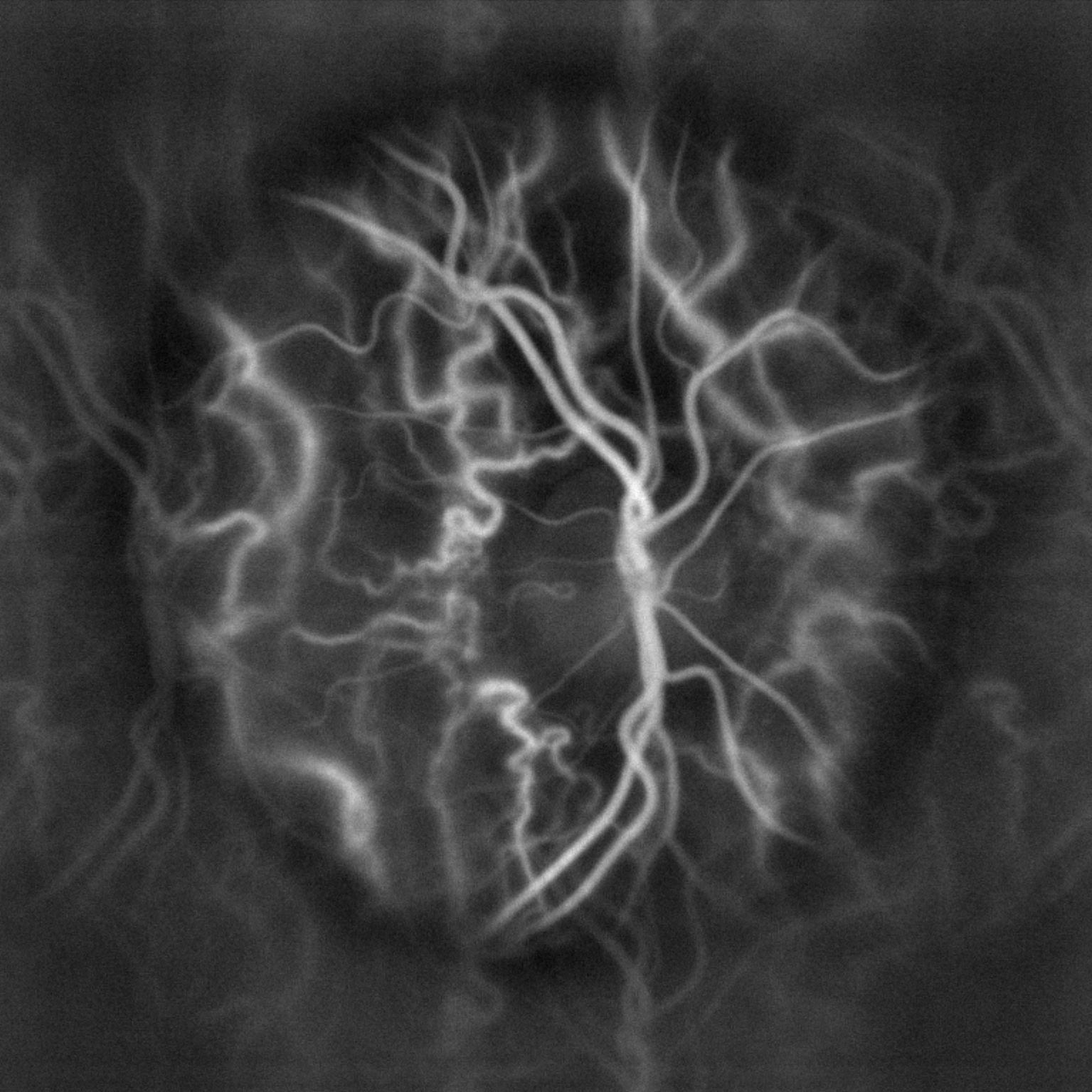}} \\
\subfigure[]{\label{eyeConfig3}\includegraphics[width=0.68\linewidth]{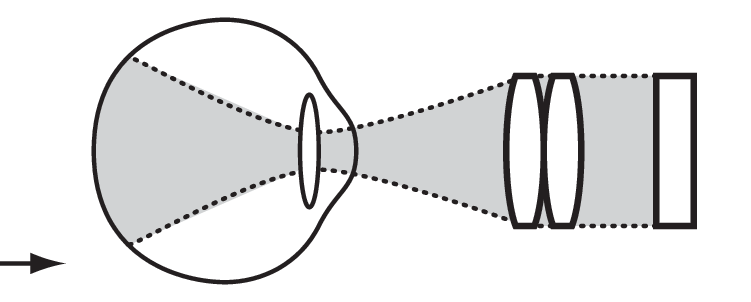}}
\subfigure[]{\label{eyeConfig3_Doppler}\includegraphics[width=0.28\linewidth,pagebox=mediabox]{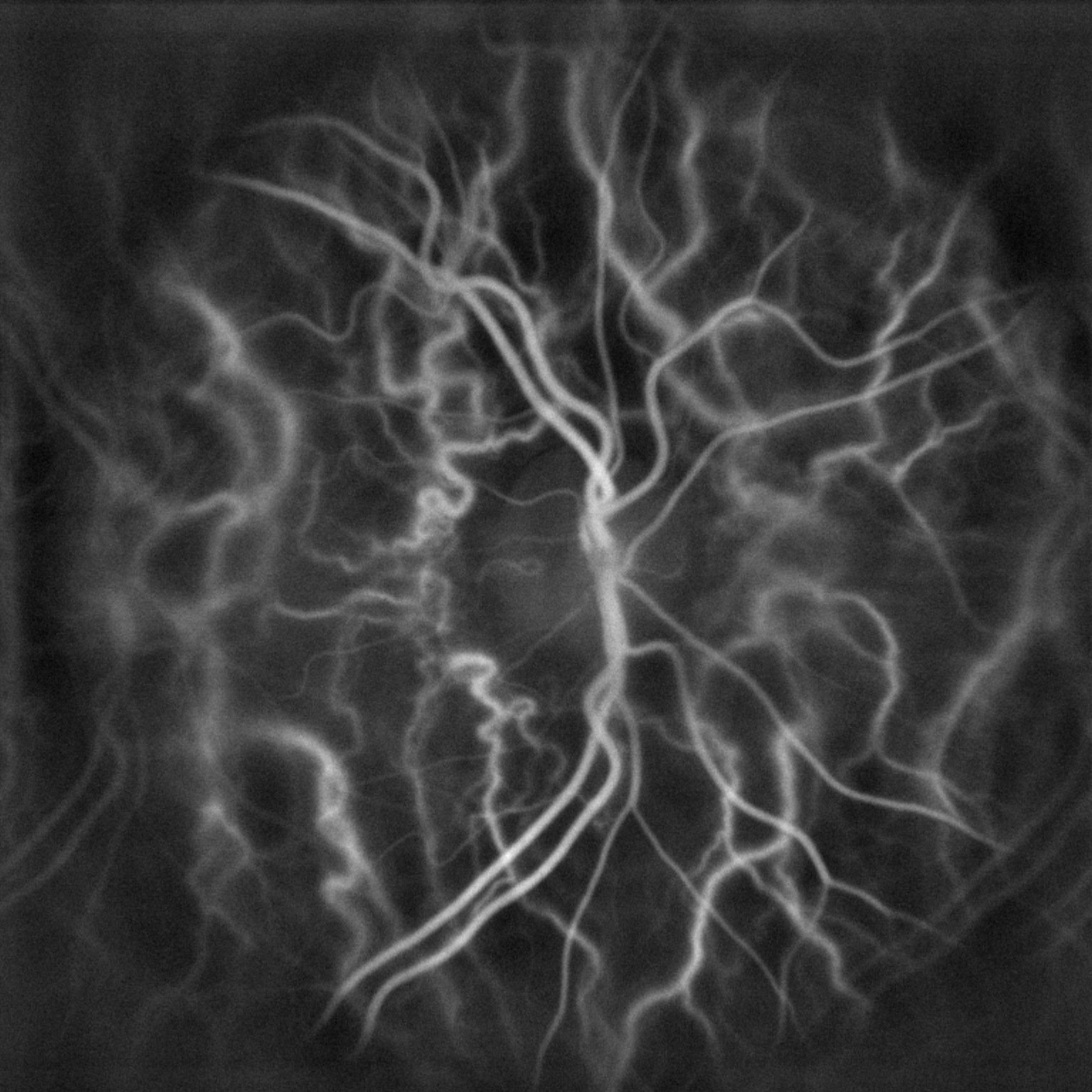}} \\
\caption{Illumination configurations and corresponding Doppler fundus images. Without diffuser, a localized hot spot is created near the laser focus in front of the cornea, and the iris acts as an image field diaphragm (a,b). With the diffuser inserted and no change in cornea--eyepiece distance, the illumination beam is broadened and made more spatially homogeneous while preserving a similar retinal field of view (c,d). With the diffuser inserted and the cornea positioned closer to the focal region of the illumination beam, the retinal field of view increases and the iris no longer limits the computed retinal field in the same way (e,f).}
\label{fig_eyeConfig}
\end{figure}

\begin{figure}[htbp]
\centering
\subfigure[]{\label{fig_Doppler1LF}\includegraphics[width=0.45\linewidth,pagebox=mediabox]{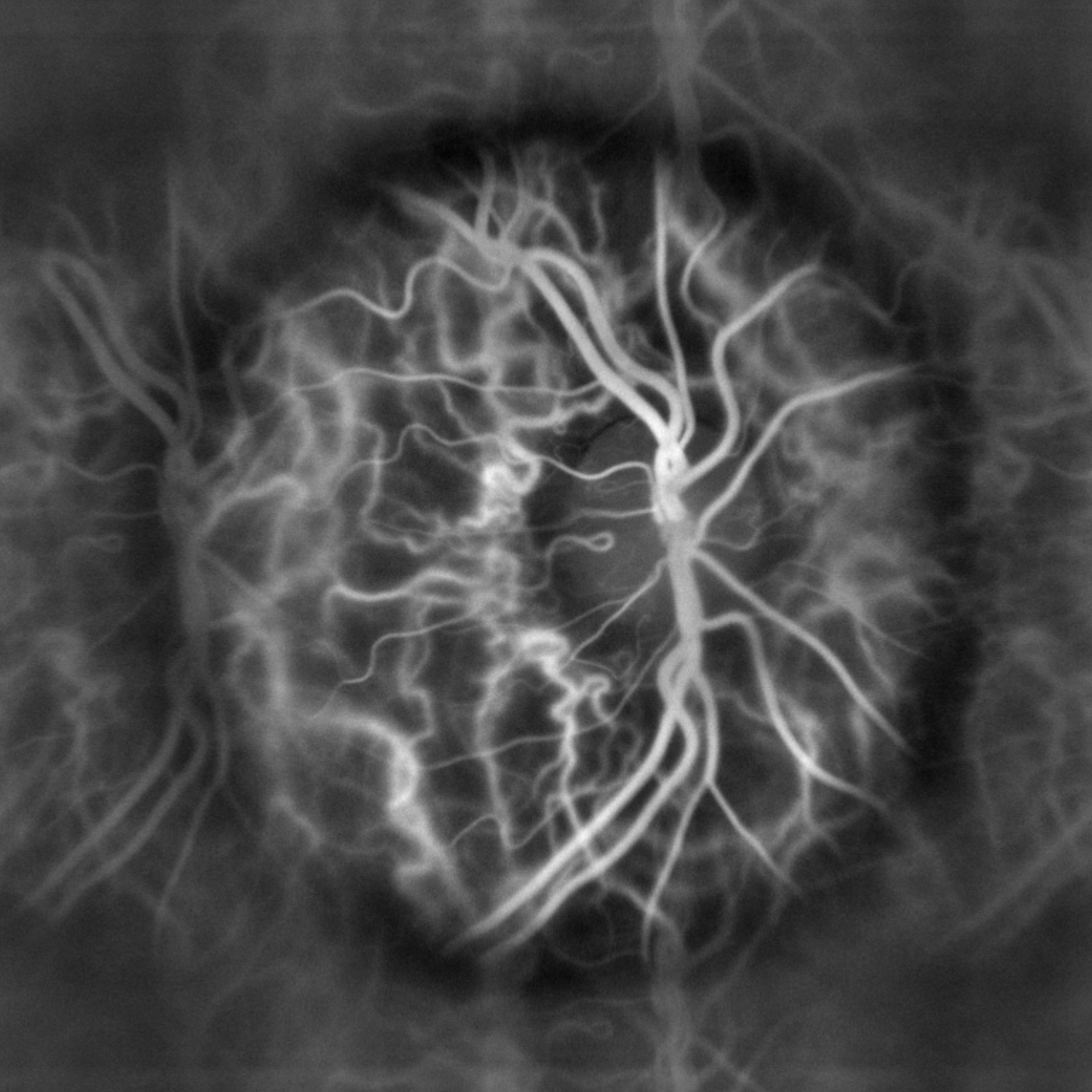}}
\subfigure[]{\label{fig_Doppler1HF}\includegraphics[width=0.45\linewidth,pagebox=mediabox]{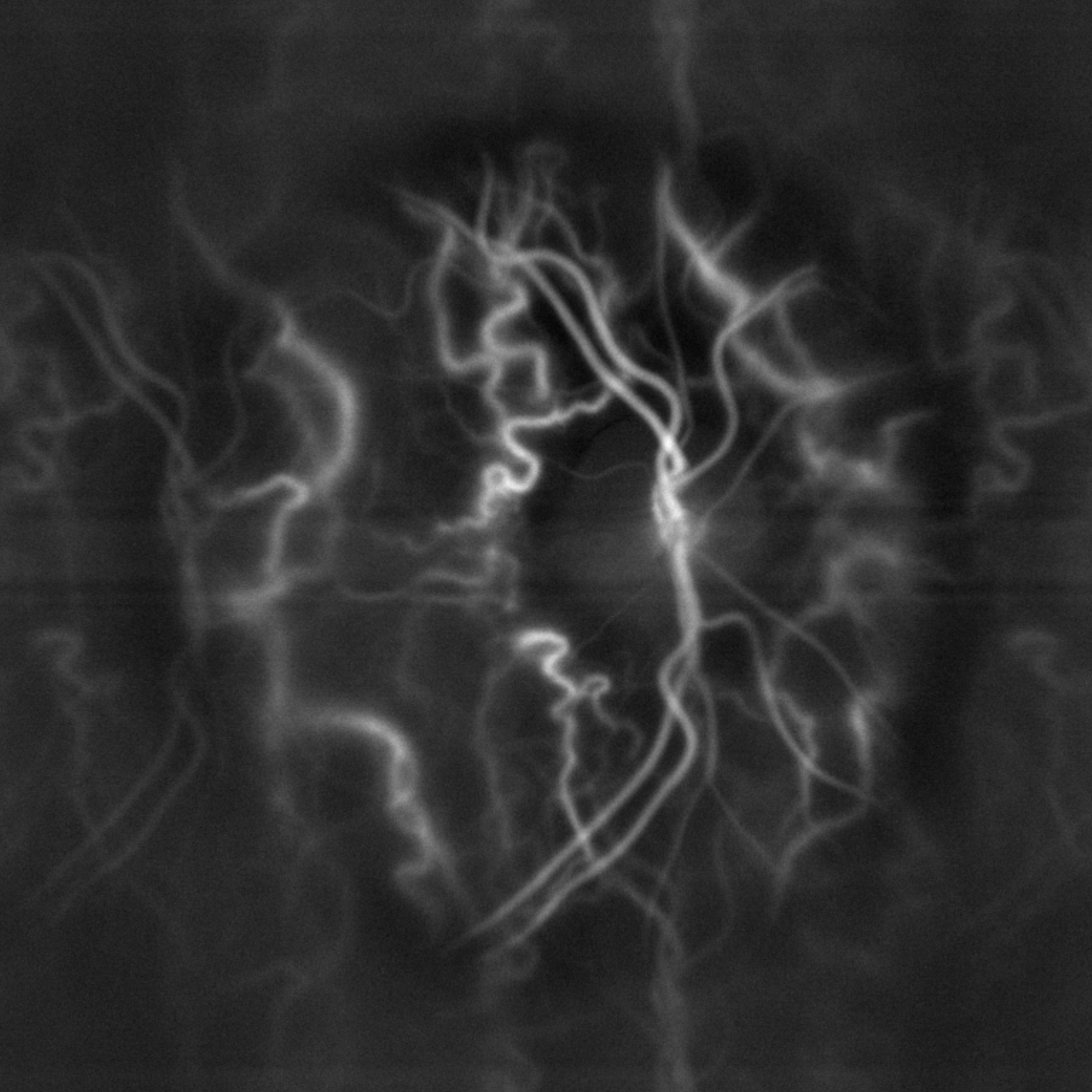}}\\
\subfigure[]{\label{fig_Doppler2LF}\includegraphics[width=0.45\linewidth,pagebox=mediabox]{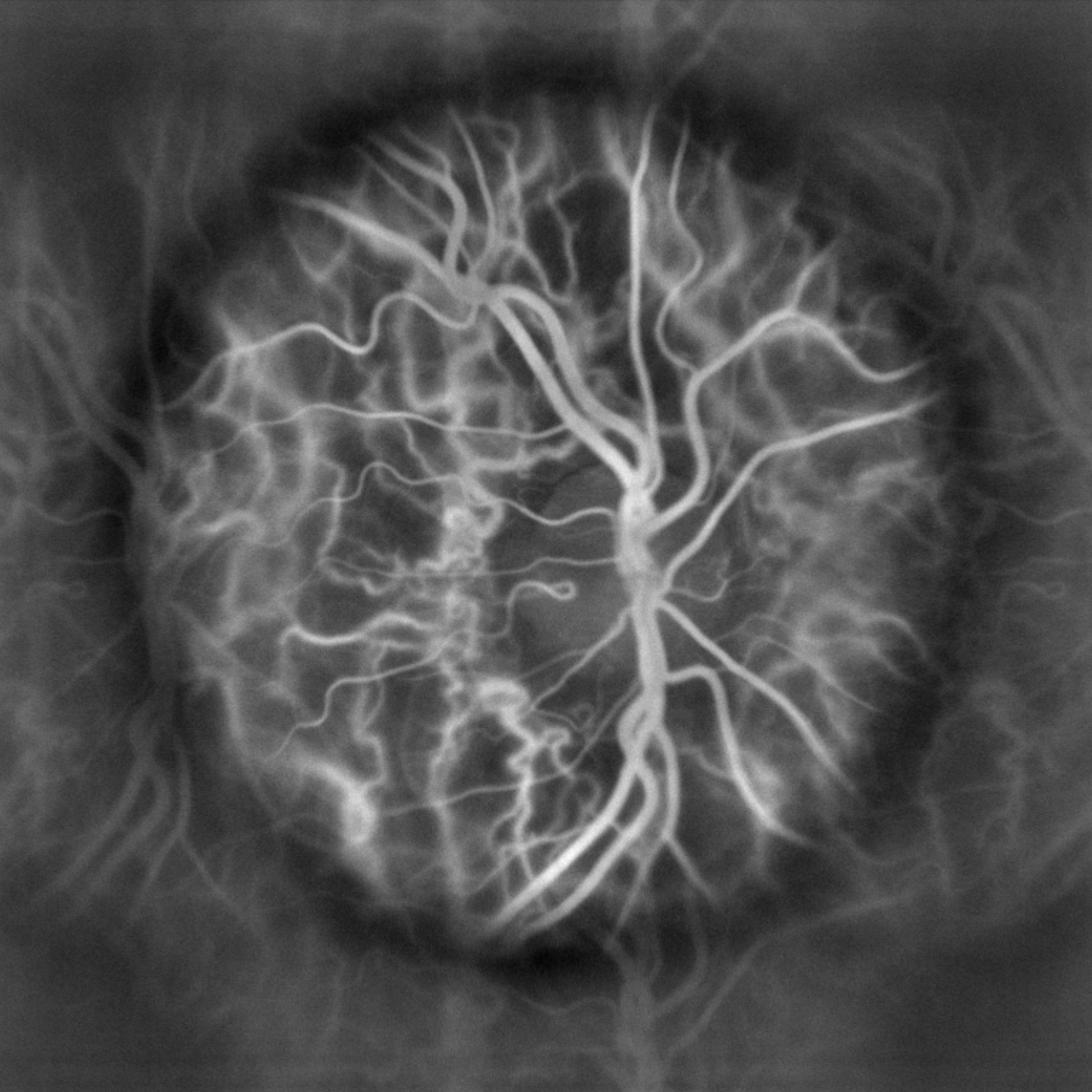}}
\subfigure[]{\label{fig_Doppler2HF}\includegraphics[width=0.45\linewidth,pagebox=mediabox]{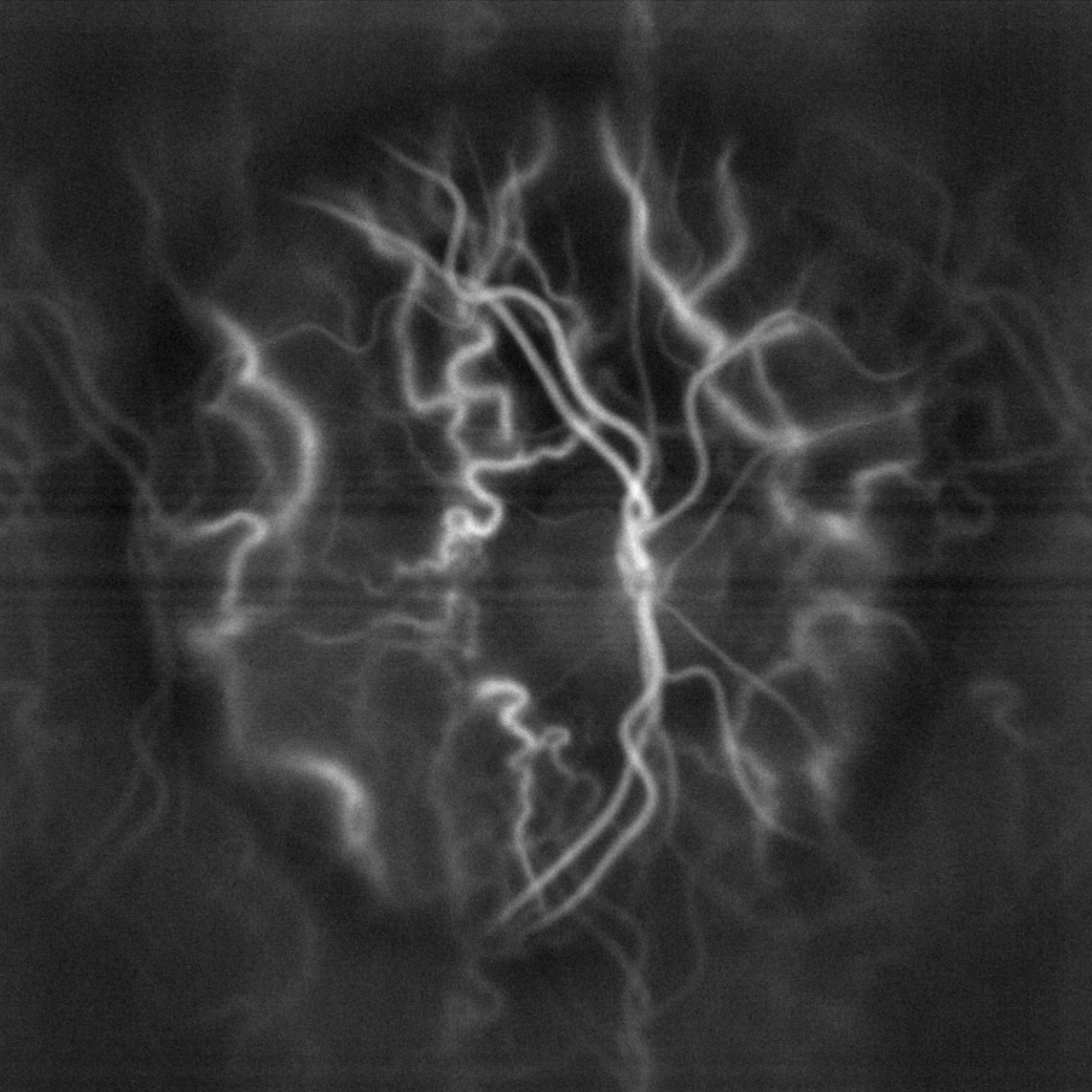}}\\
\subfigure[]{\label{fig_Doppler3LF}\includegraphics[width=0.45\linewidth,pagebox=mediabox]{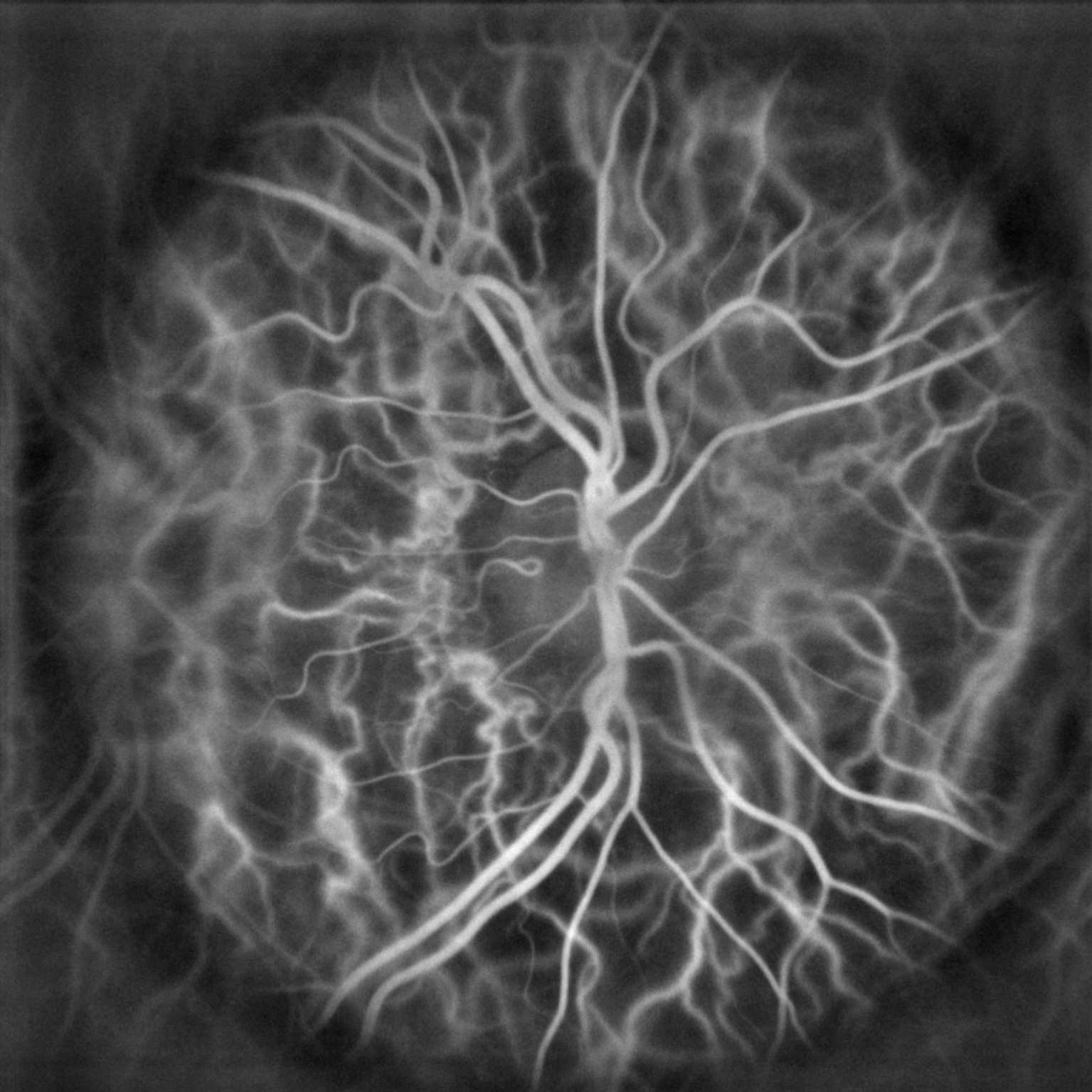}}
\subfigure[]{\label{fig_Doppler3HF}\includegraphics[width=0.45\linewidth,pagebox=mediabox]{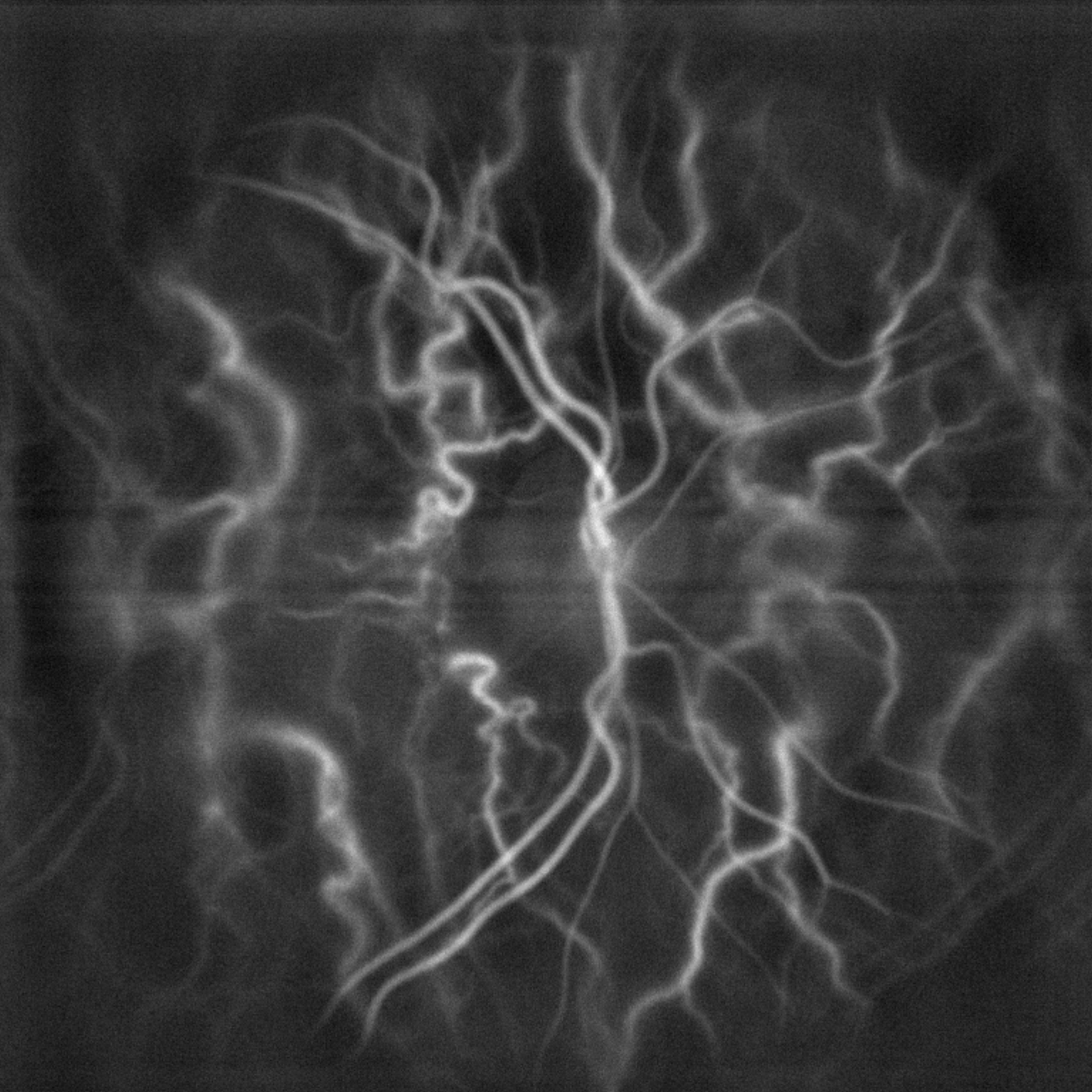}}
\caption{Doppler fundus images from the broad fluctuation frequency band between 2 kHz and 17 kHz (left column) and from the high-frequency band between 12 kHz and 17 kHz (right column). Images were computed from raw interferograms acquired in the three layouts shown in Fig.~\ref{fig_eyeConfig}. Top row: focused non-diffuse illumination. Middle row: diffuse illumination with the same cornea--eyepiece distance as in the non-diffuse case. Bottom row: diffuse Maxwellian illumination with reduced cornea--eyepiece distance to increase the retinal field of view.}
\label{fig_DopplerImagesVsConfig}
\end{figure}

The experimental setup is based on an inline Mach--Zehnder interferometer (Fig.~\ref{fig_Setup}). The near-infrared radiation from a diode laser (Thorlabs FPV852P, wavelength $\lambda=852$ nm, model 40750) is split 10\%--90\% into linearly polarized reference and illumination arms, respectively, emerging from polarization-maintaining fibers (Thorlabs PM780-HP, numerical aperture NA $\sim0.12$). The illumination beam passes through an engineered diffuser (Thorlabs ED1-C20-MD SM1-threaded mount, diameter 1 inch, 20$^\circ$ circular top-hat engineered diffuser) and an eyepiece made of two biconvex lenses of 60 mm focal length each, with an effective focal length of $\sim$33 mm.

Two cameras record interferograms of the cross-polarized backscattered light component with respect to illumination. One camera is used for real-time preview (Adimec Quartz Q-2HFW-Hm/CXP-6-0.5, pixel pitch 12 $\mu$m). A high-speed camera is used for offline image rendering (Ametek Phantom V-series camera, frame rate 35 kHz, pixel pitch $d=28~\mu$m, frame size $N_x\times N_y=768\times768$ pixels). The latter is used for the interferogram measurements used in image rendering.

The retina of a volunteer was illuminated with a continuous-wave laser beam focused through the eyepiece in three layouts, sketched in Fig.~\ref{fig_eyeConfig}:
\begin{itemize}
\item the illumination beam was focused in front of the eye in the absence of diffuser (Fig.~\ref{eyeConfig1});
\item the diffuser was introduced while keeping the same relative position between the cornea and the eyepiece (Fig.~\ref{eyeConfig2});
\item the diffuser was kept in place and the cornea--eyepiece distance was reduced to increase the retinal field of view (Fig.~\ref{eyeConfig3}).
\end{itemize}

Informed consent was obtained from the subject. Experimental procedures adhered to the tenets of the Declaration of Helsinki. Study authorization was obtained from the appropriate local ethics review boards: Personal Protection Committees (CPP Sud-Est III: 2019-021B) and the National Agency for the Safety of Medicines and Health Products (ANSM No. IRDCB: 2019-A00942-55). The clinical trial was registered under reference NCT04129021.

Patient positioning was monitored by real-time computation and visualization of clutter-free inline digital holograms of the eye fundus from an input stream of 16-bit, 1024-by-1024-pixel interferograms recorded at 800 frames per second with the Adimec Quartz camera. This was done by Fresnel transformation and principal component analysis of stacks of 64 consecutive holograms \cite{PCA2020}, with the digital hologram streaming software \href{http://holovibes.com/}{HoloVibes}.

\section{Optical configuration and interference-pattern measurement}

\begin{figure}[htbp]
\centering
\includegraphics[width=\linewidth]{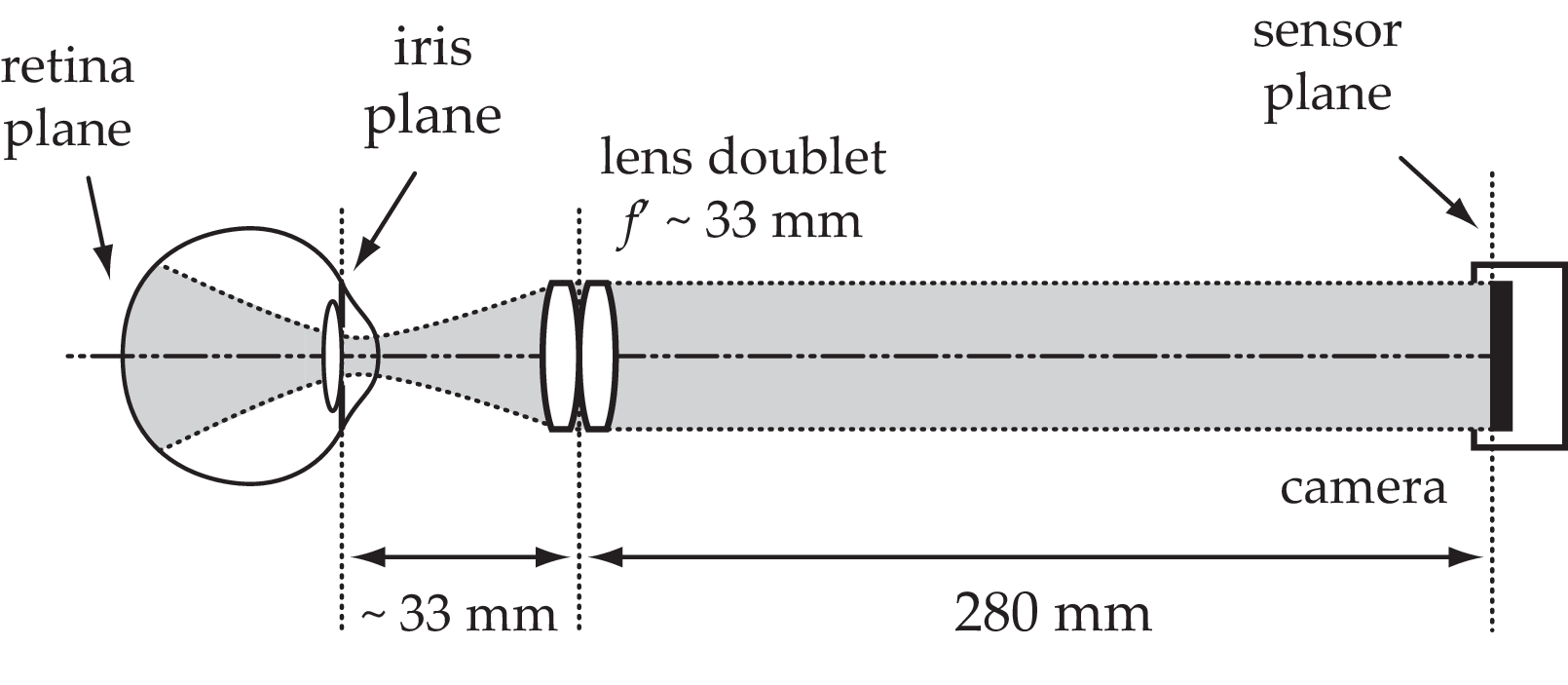}
\caption{Illustrative optical configuration for Maxwellian-view, wide-field holographic imaging of the eye fundus with diffuse laser illumination.}
\label{fig_eyeConfigConjugations}
\end{figure}

Figure~\ref{fig_eyeConfigConjugations} illustrates the optical configuration used for interference-pattern measurement and digital image rendering. Each eyepiece lens has a focal length of 60 mm (Thorlabs LB1723-B; N-BK7 biconvex lens, diameter 2 inches, $f=60.0$ mm, antireflection coating 650--1050 nm). The doublet has an effective focal length $f'\sim33$ mm for a distance between its composing lenses of approximately 1 cm, according to Gullstrand's formula.

The optical conjugate of the sensor plane is set in front of the cornea, at a distance $\sim33$ mm from the center of the doublet. At this distance, the sensor-to-cornea magnification ratio is $M=33/280\sim0.118$. To estimate the sensor-to-retina magnification ratio of the digitally rendered image, the optical system of the eye is considered in the thin-lens approximation. For an average axial length of the eye of 25 mm, $M'=(33+25)/280\sim0.207$. The pixel pitch of the rendered image with the angular spectrum propagation method, for numerical reconstruction distances $z<N_xd^2/\lambda\simeq0.71~\rm m$ from the cornea to the retina, has a constant value for a given magnification ratio.

The image pitch in the iris and retina planes are therefore $Md\sim3.3~\mu\rm m$ and $M'd\sim5.8~\mu\rm m$, respectively. The lateral fields of view of the rendered image in the iris and retina planes are $N_xMd\sim2.5~\rm mm$ and $N_xM'd\sim4.4~\rm mm$, respectively. The former typically fits within the normal adult pupil diameter, which varies from approximately 2--4 mm in bright light to 4--8 mm in the dark \cite{Spector1990}. Thus, the lateral field at the iris plane is of the order of the iris aperture, while the digitally reconstructed retinal field is increased. This configuration can be adapted by changing the cornea--eyepiece distance, the eyepiece focal length, or the sensor--eyepiece distance.

\section{Digital image rendering}

Offline computation and registration of Doppler images from 12-bit, 768-by-768-pixel interferograms recorded at 35,000 frames per second were performed by angular-spectrum propagation \cite{Puyo2018}, singular-value-decomposition filtering, and short-time Fourier transformation \cite{Puyo2020}, with the image-rendering software \href{http://www.holodoppler.org/}{HoloDoppler}. Doppler fundus images were rendered from the broad fluctuation frequency band between 2 kHz and 17 kHz and from the high-frequency band between 12 kHz and 17 kHz. Representative images are displayed in Fig.~\ref{fig_DopplerImagesVsConfig}. These images were computed from raw interferograms acquired in the three layouts of Fig.~\ref{fig_eyeConfig}.

\section{Radiometric measurements and irradiance estimation}

Optical exposure to the eye is governed by irradiance and radiant-exposure limits applied to defined anatomical planes and averaging apertures. To document the illumination geometry, we measured the transverse intensity distributions, beam profiles, and total optical power of the near-infrared illumination beam.

\begin{figure}[htbp]
\centering
\subfigure[]{\label{hotspot1}\includegraphics[width=0.4\linewidth,pagebox=mediabox]{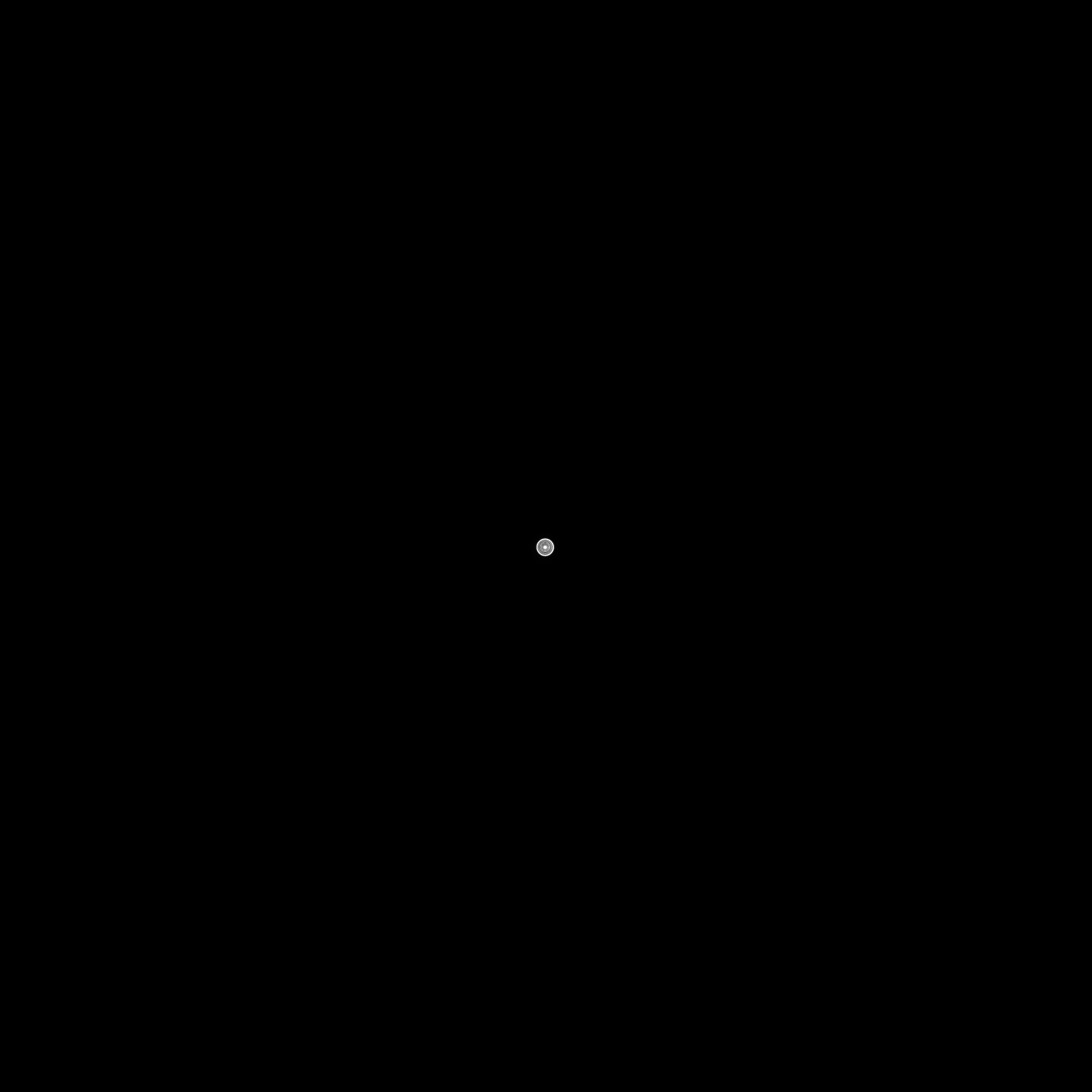}}
\subfigure[]{\label{hotspot2}\includegraphics[width=0.4\linewidth,pagebox=mediabox]{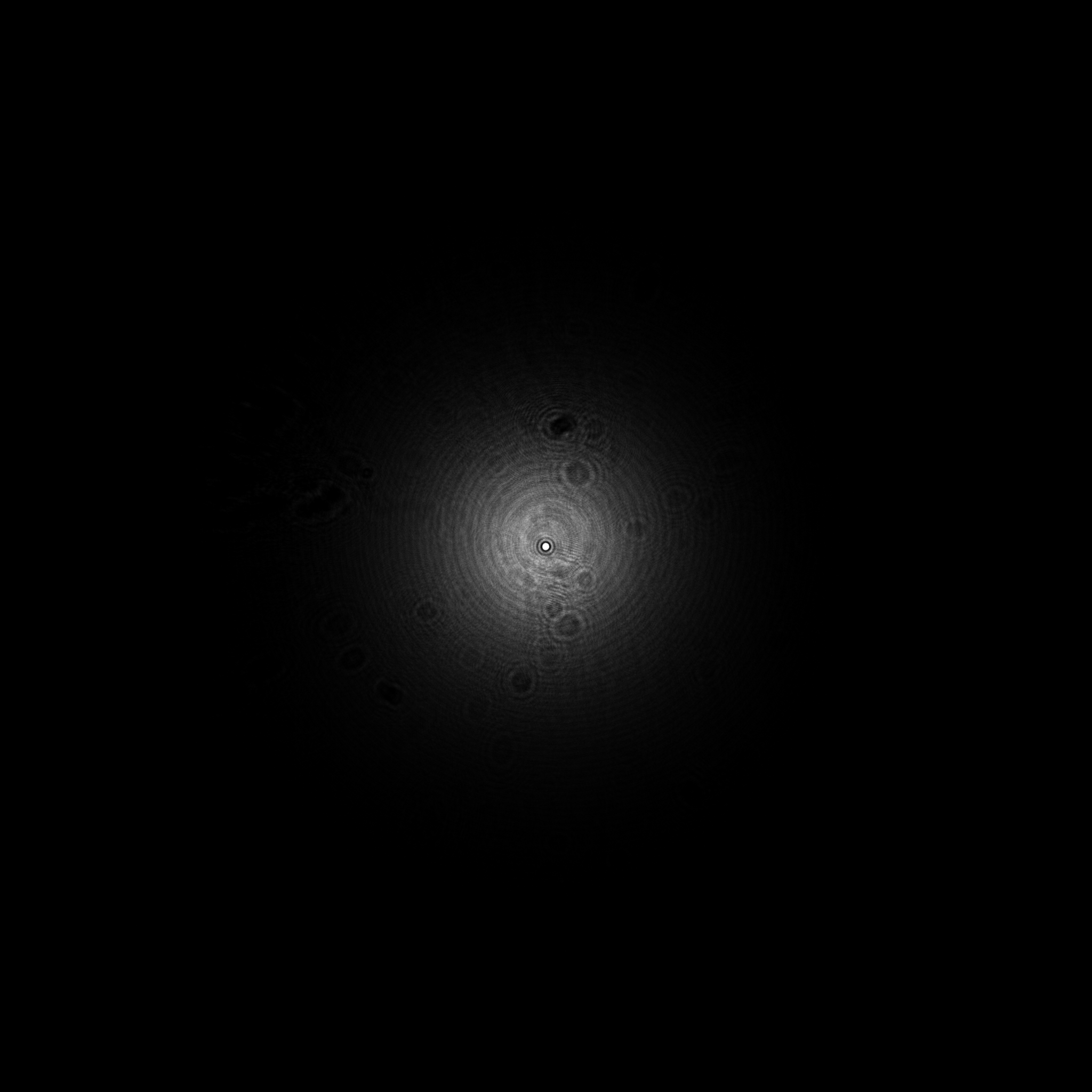}}\\
\subfigure[]{\label{hotspot3}\includegraphics[width=0.4\linewidth,pagebox=mediabox]{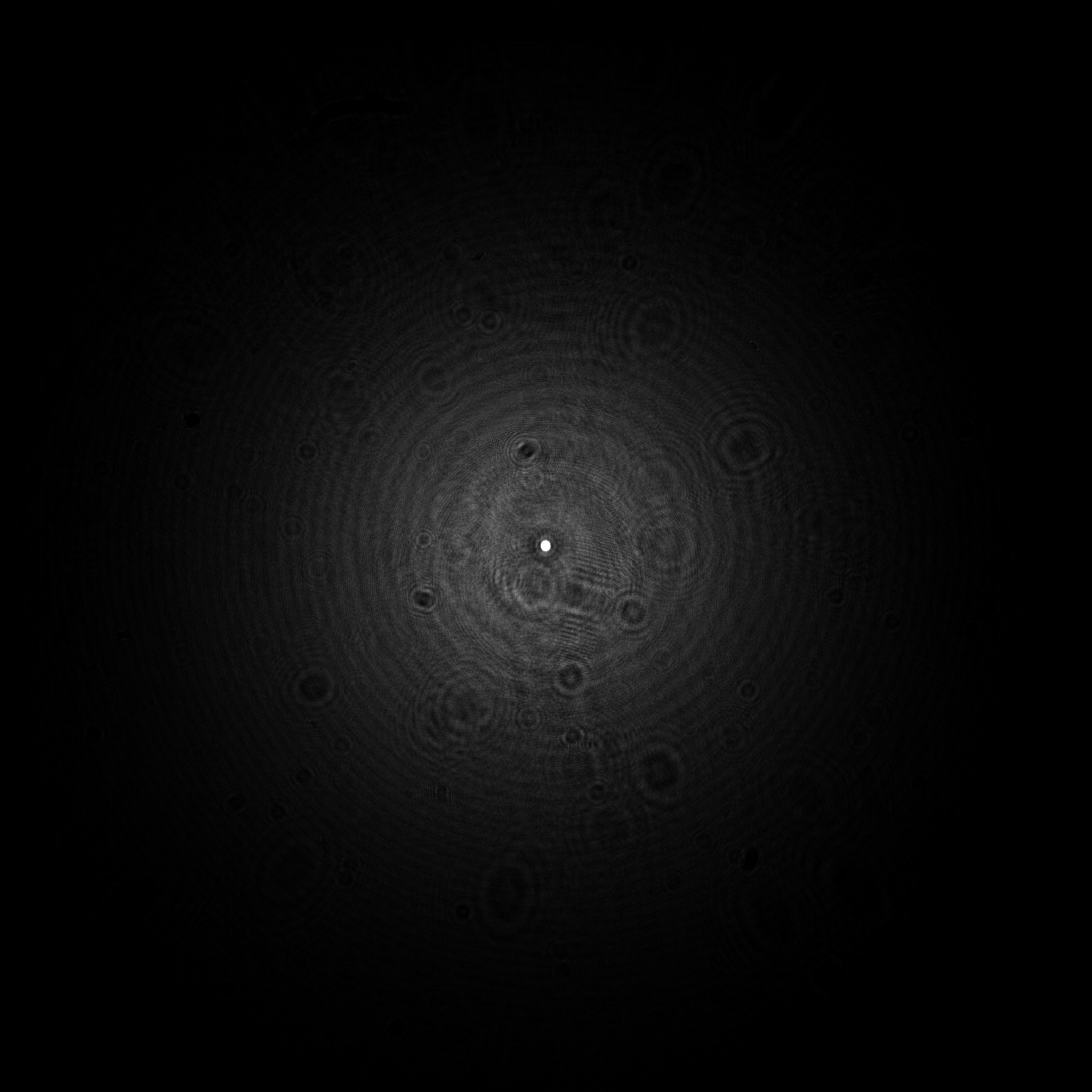}}
\subfigure[]{\label{hotspot4}\includegraphics[width=0.4\linewidth,pagebox=mediabox]{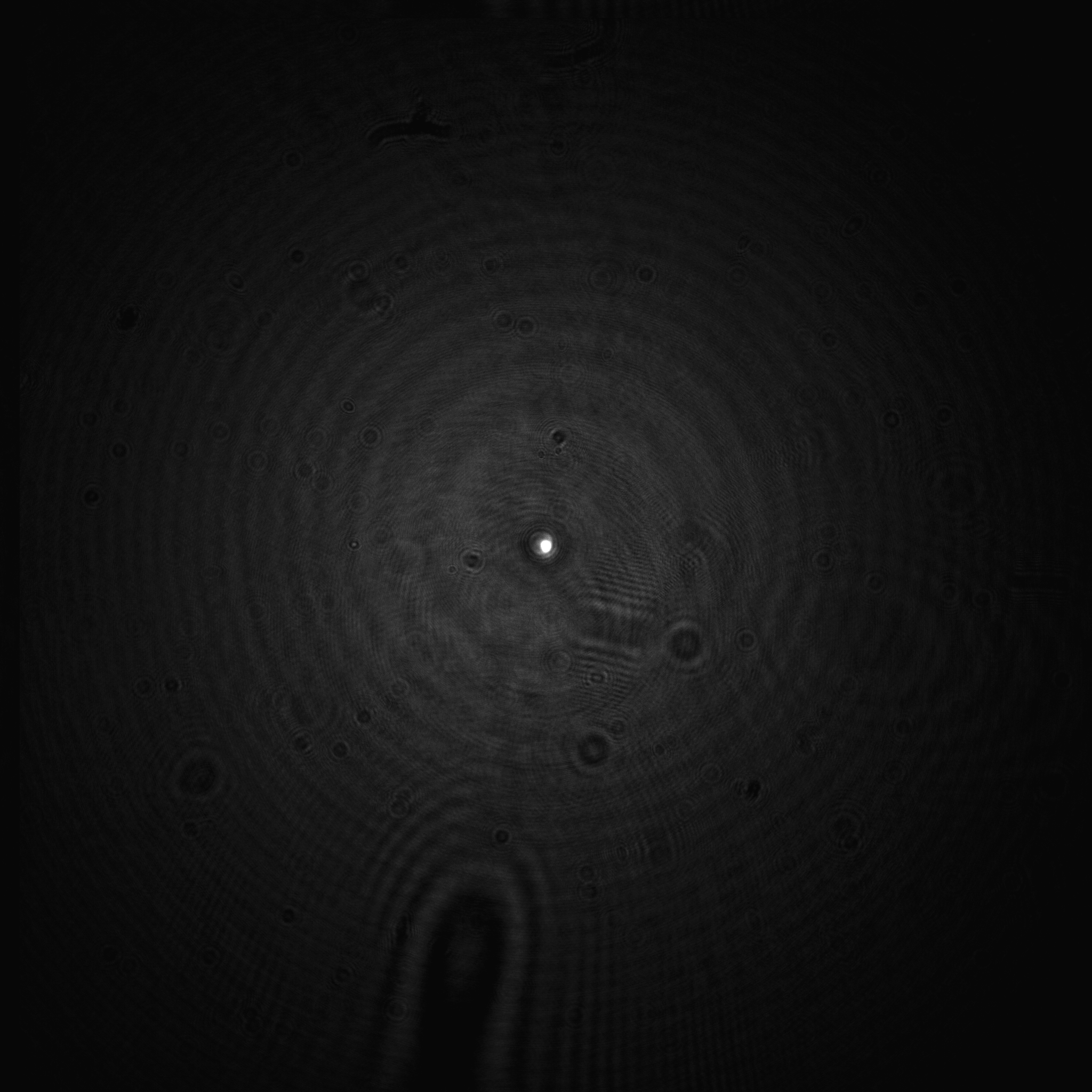}}\\
\subfigure[]{\label{Config1}\includegraphics[width=0.7\linewidth]{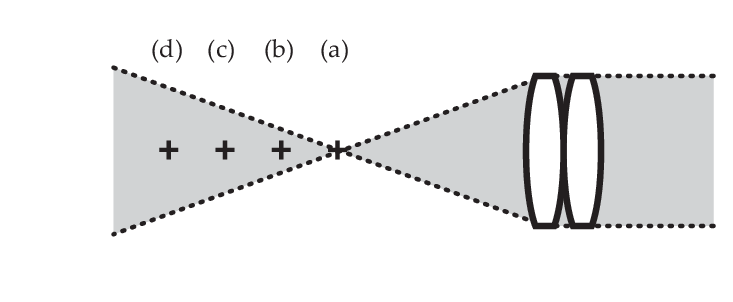}}\\
\subfigure[]{\label{crosssectionprofilehotspot}\includegraphics[width=0.8\linewidth]{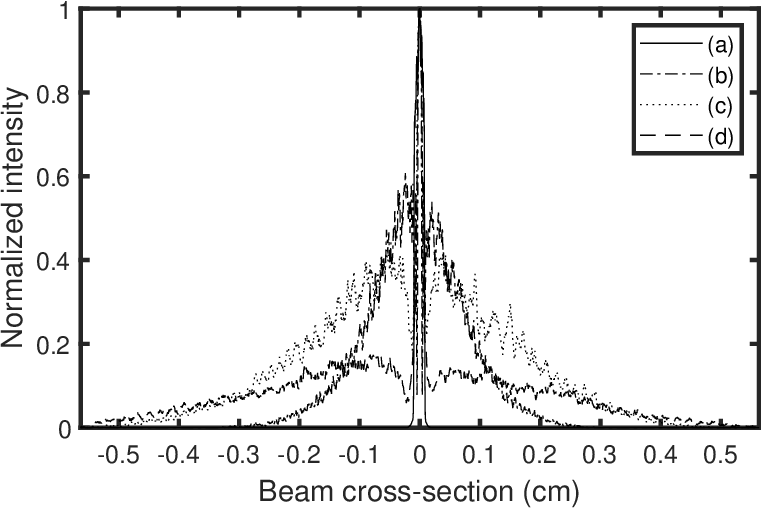}}
\caption{Intensity distribution at different planes from the illumination beam waist without diffuser, toward the eye: 0.0 mm \subref{hotspot1}, 7.5 mm \subref{hotspot2}, 15 mm \subref{hotspot3}, and 22.5 mm \subref{hotspot4}. The measurement locations are sketched in \subref{Config1}; radial cross-section profiles are shown in \subref{crosssectionprofilehotspot}.}
\label{fig_nodiffdistrib}
\end{figure}

\begin{figure}[htbp]
\centering
\subfigure[]{\label{csdiff1}\includegraphics[width=0.4\linewidth,pagebox=mediabox]{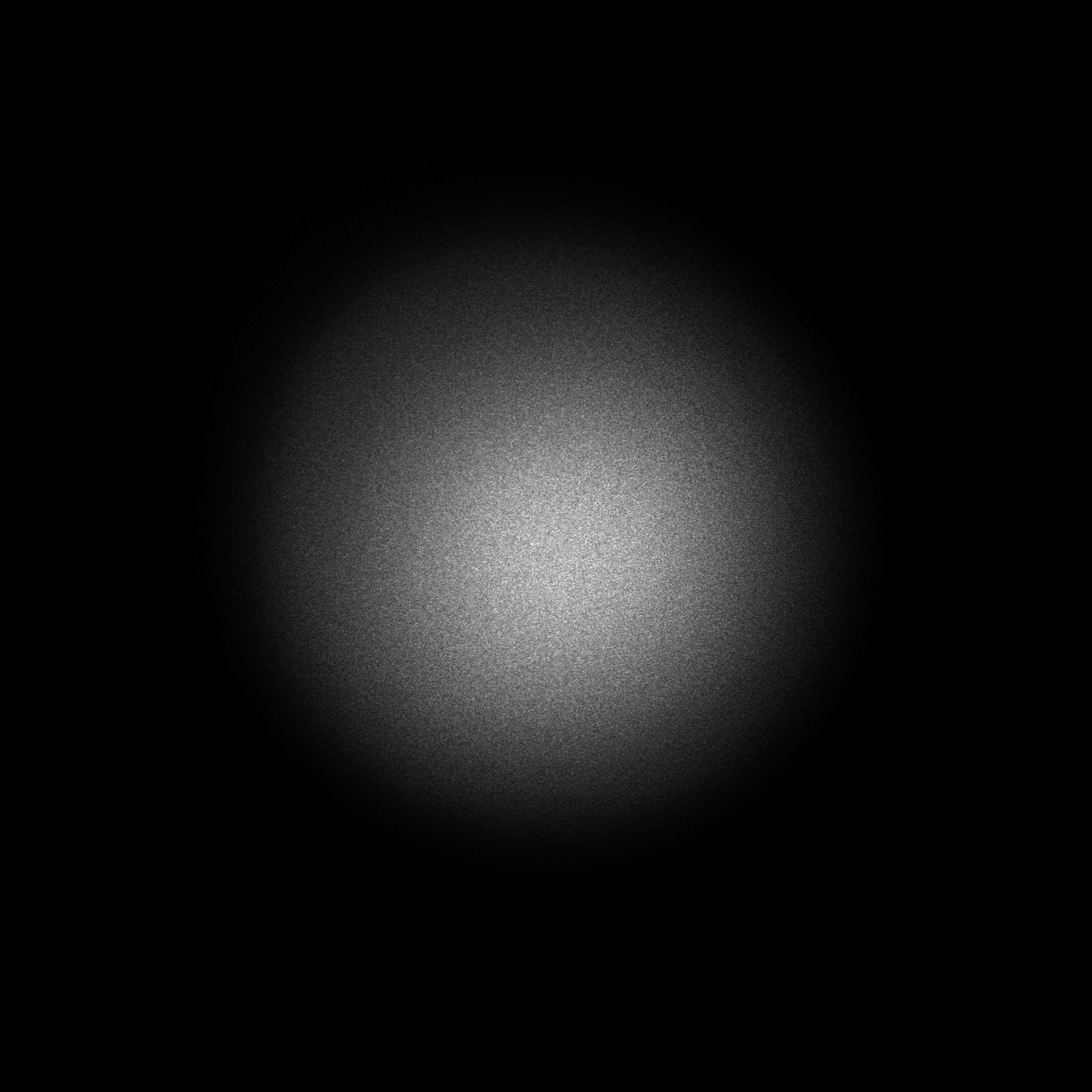}}
\subfigure[]{\label{csdiff2}\includegraphics[width=0.4\linewidth,pagebox=mediabox]{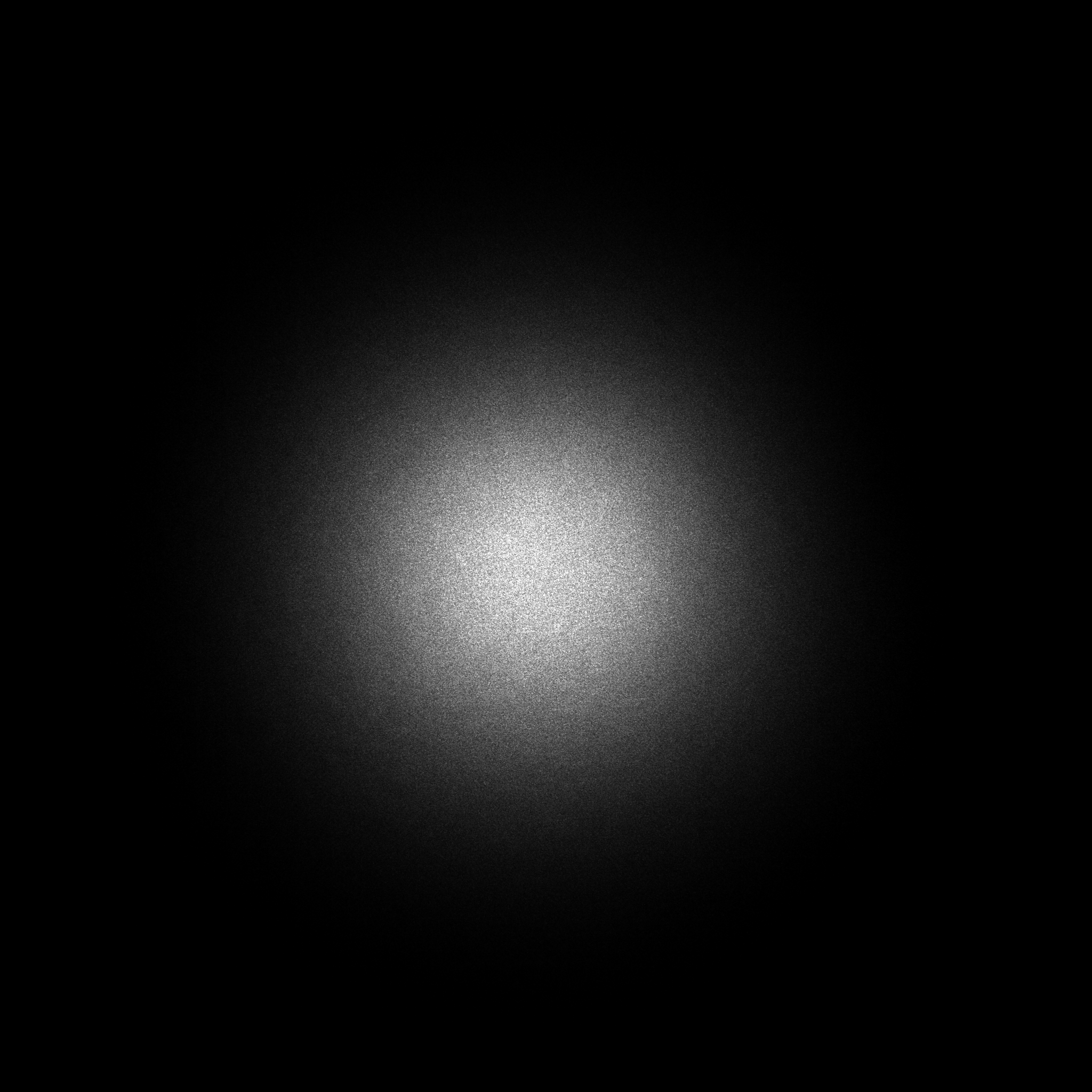}}\\
\subfigure[]{\label{csdiff3}\includegraphics[width=0.4\linewidth,pagebox=mediabox]{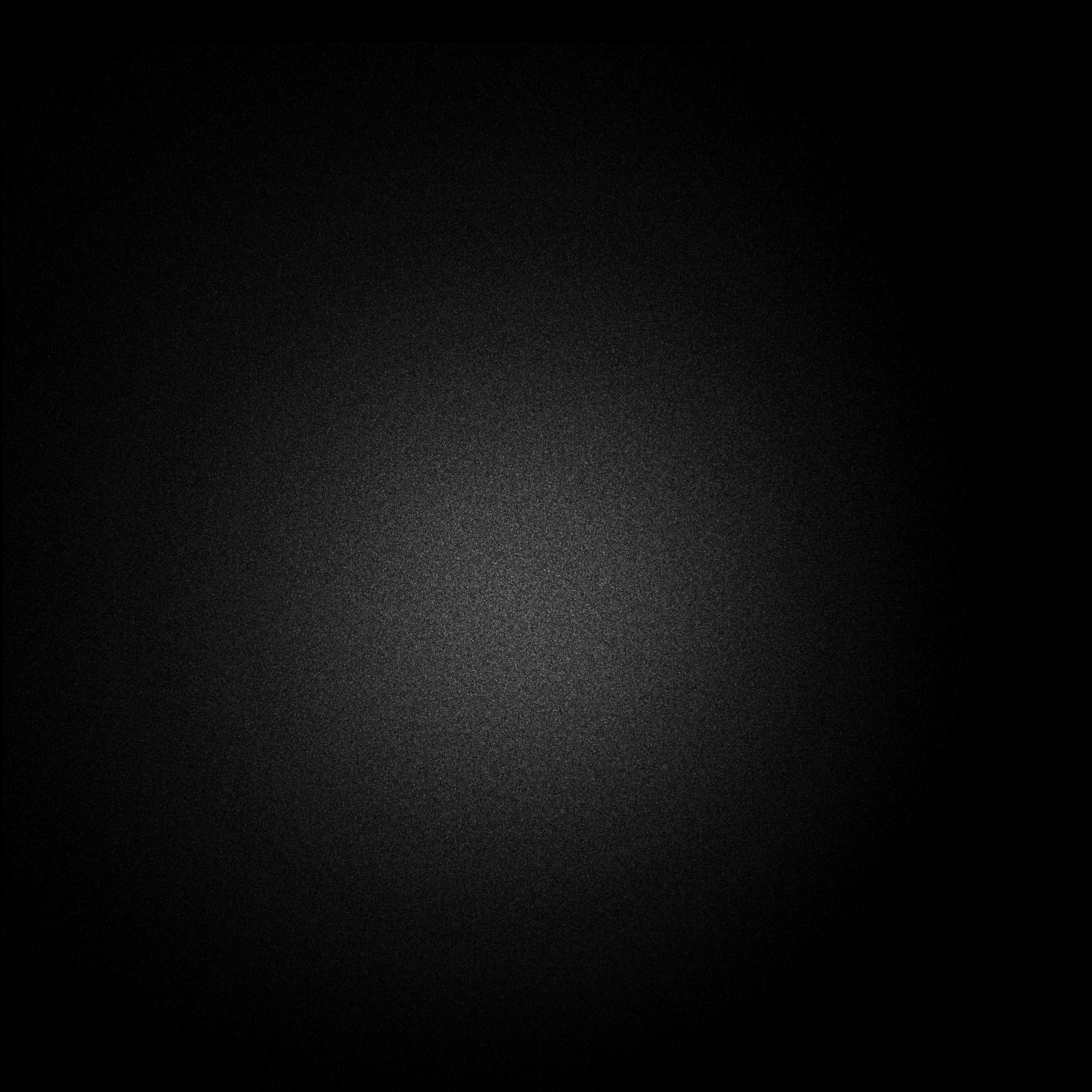}}
\subfigure[]{\label{csdiff4}\includegraphics[width=0.4\linewidth,pagebox=mediabox]{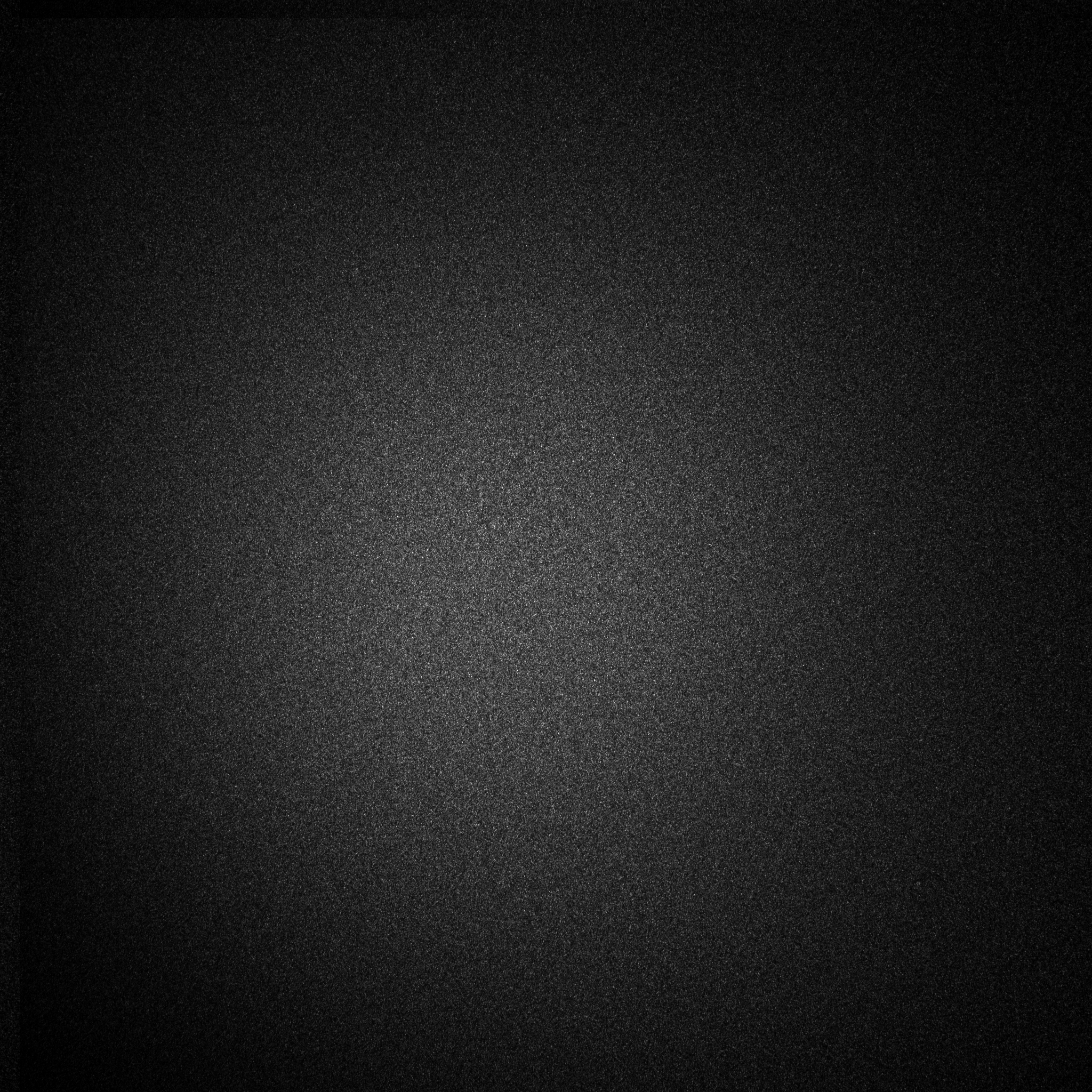}}\\
\subfigure[]{\label{Config3}\includegraphics[width=0.7\linewidth]{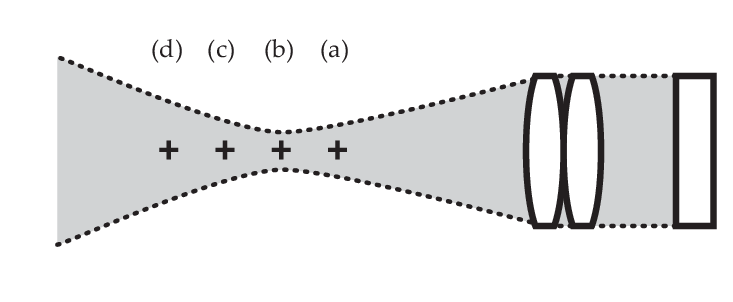}}\\
\subfigure[]{\label{crosssectionprofilediff}\includegraphics[width=0.8\linewidth]{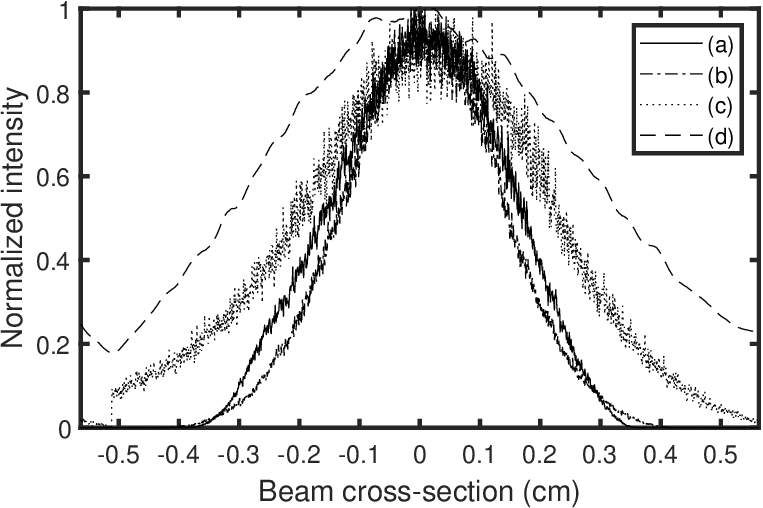}}
\caption{Intensity distribution at different planes from the non-diffuse beam waist, after insertion of the diffuser, toward the eye: 0.0 mm \subref{csdiff1}, 7.5 mm \subref{csdiff2}, 15 mm \subref{csdiff3}, and 22.5 mm \subref{csdiff4}. The measurement locations are sketched in \subref{Config3}; radial cross-section profiles are shown in \subref{crosssectionprofilediff}.}
\label{fig_diffdistrib}
\end{figure}

\subsection{Optical power}

The optical power of the illumination beam was measured with a photodiode power sensor and digital console (Thorlabs S121C and PM100D). The beam power was measured in the waist plane over an active detection area of 9.7 mm $\times$ 9.7 mm with an entrance-aperture diameter of 9.5 mm.

\subsection{Beam cross-section measurement}
\label{sect_beamXsectMeasurement}

The irradiance distributions at planes relevant to corneal positioning were evaluated with and without the diffuser (Figs.~\ref{fig_diffdistrib} and \ref{fig_nodiffdistrib}). Measurements were taken at different distances from the focal plane using the bare sensor array of a camera (XIMEA XIQ MQ042xG-CM, 2048-by-2048 pixels, pixel pitch 5.5 $\mu$m). The horizontal cross-section profiles in Figs.~\ref{crosssectionprofilehotspot} and \ref{crosssectionprofilediff}, averaged over approximately 165 $\mu$m vertically around the center of the distribution, describe the radial distribution of light intensity near the corneal plane.

\begin{figure}[htbp]
\centering
\subfigure[]{\label{IrradianceMap}\includegraphics[width=0.8\linewidth]{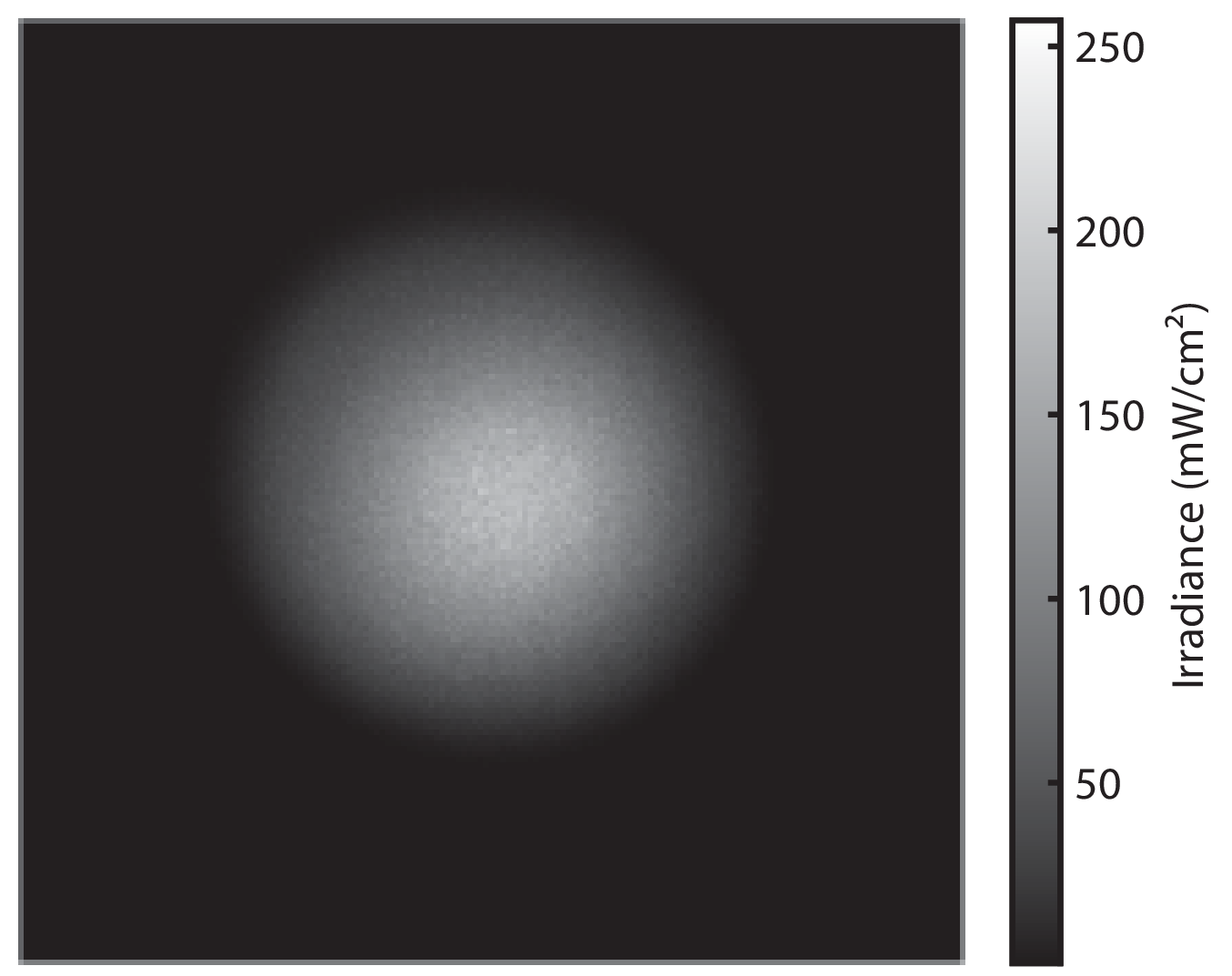}}\\
\subfigure[]{\label{IrradianceCutH}\includegraphics[width=0.8\linewidth]{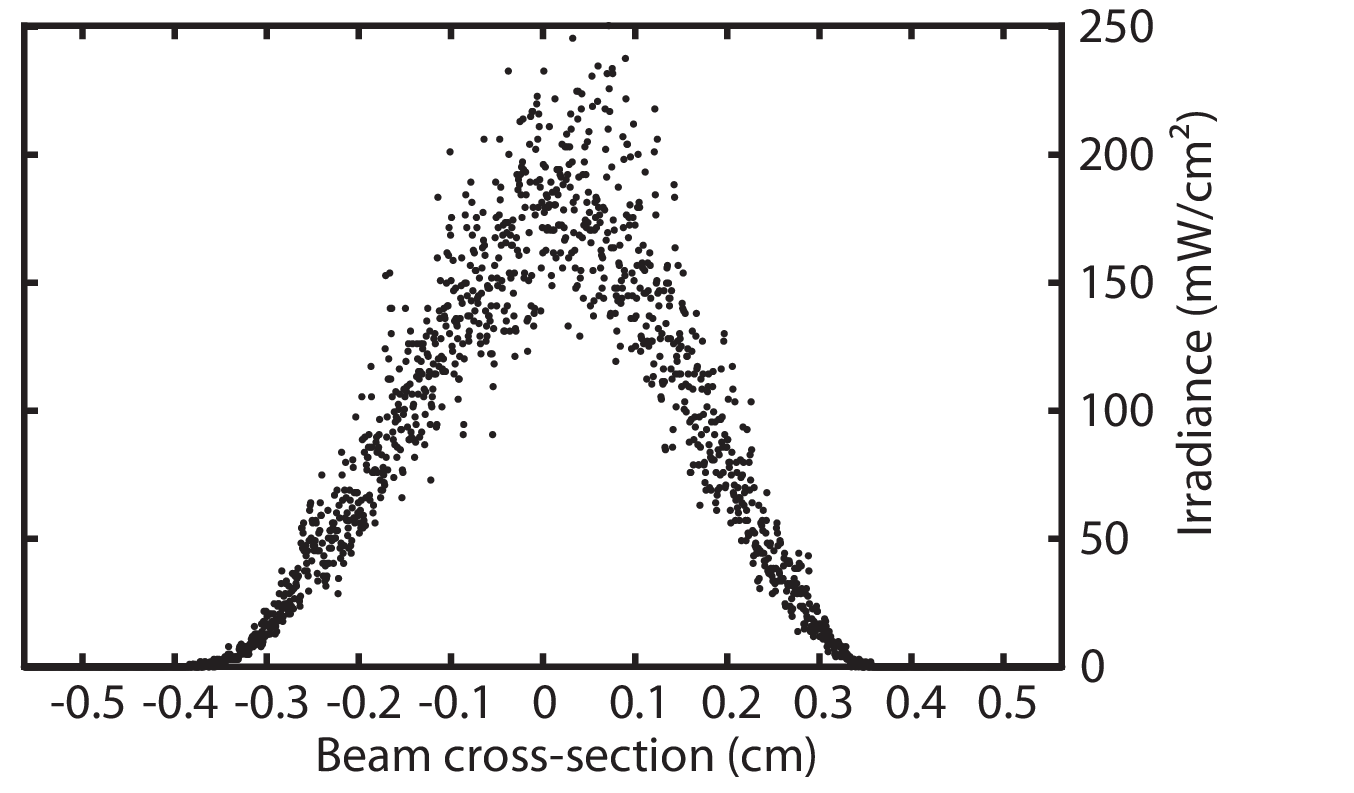}}
\caption{Irradiance map \subref{IrradianceMap} and mid-horizontal cross-section \subref{IrradianceCutH} assessed at the waist of the diffuse illumination beam, onto each camera pixel, for a total optical power of 22 mW.}
\label{fig_maxIrradiance_retina}
\end{figure}

A quantitative irradiance map was assessed at the waist of the diffuse illumination beam from the measured distribution and scaled to a total illumination power of 22 mW (Fig.~\ref{fig_maxIrradiance_retina}). This scaling provides a reference for estimating the maximum local irradiance at other operating powers by linear rescaling.

\section{Light-hazard assessment according to current ophthalmic standards}
\label{sec:current_standards}

The light-hazard analysis was restricted to two current ophthalmic standards: ISO 15004-2:2024, for the European framework, and ANSI Z80.36-2021, for the American framework. Both standards are written for ophthalmic instruments that direct optical radiation into or at the eye for diagnostic, imaging, measurement, or alignment purposes. The analysis below concerns the described research configuration and does not constitute regulatory certification of a medical device. Final conformity assessment must be performed for the final intended-use configuration, including wavelength, exposure duration, optical power, diffuser, eyepiece geometry, acquisition protocol, patient-positioning procedure, and all simultaneously or consecutively used optical sources.

The illumination source used here is a continuous-wave, monochromatic near-infrared laser at $\lambda=852$ nm. For this wavelength, the relevant hazards are visible/infrared retinal thermal exposure, iris exposure when the iris is intentionally or incidentally illuminated, and anterior-segment exposure if the beam diameter becomes sufficiently small between the cornea and the posterior crystalline lens. Ultraviolet and blue-light photochemical criteria are not limiting for this 852 nm source.

For a conservative design calculation, the measured diffuse beam distribution is treated as the narrowest relevant illumination distribution in the Maxwellian-view configuration. The irradiance map in Fig.~\ref{fig_maxIrradiance_retina}, scaled to 22 mW total power, has a maximum local pixel irradiance of approximately $252~\rm mW/cm^2$. Because irradiance scales linearly with optical power, the peak-pixel irradiance at any total power $P$ is estimated as
\begin{equation}
E_{\rm pix,max}(P) \simeq 252\,\frac{P}{22~\rm mW}\;\rm mW/cm^2.
\label{eq:peak_pixel_scaling}
\end{equation}
This peak-pixel value is conservative relative to exposure quantities averaged over regulatory apertures larger than one camera pixel. The final assessment should nevertheless be based on the moving-aperture average prescribed by the applicable standard. In practice, this is obtained by convolving the measured irradiance map with the relevant circular averaging aperture and retaining the maximum value across the field.

\subsection{ISO 15004-2:2024 assessment}
\label{sec:iso2024_assessment}

ISO 15004-2:2024 classifies ophthalmic instruments into Group 1 and Group 2 and provides exposure limits for Group 1 classification. For the present continuous-wave 852 nm illumination, the relevant Group 1 quantities are the corneal/lenticular visible and infrared exposure, iris visible and infrared thermal exposure, and retinal visible and infrared thermal exposure.

For large retinal spots, ISO 15004-2:2024 gives a retinal visible/infrared thermal radiant-exposure limit that, for exposure durations $0.25~{\rm s}<t\leq2.5\times10^3~{\rm s}$, can be written as
\begin{equation}
H_{\rm VIR-R,lim}(t)=5.0\,t^{3/4}\;\rm J/cm^2,
\label{eq:iso_retina_H}
\end{equation}
and therefore as an equivalent average irradiance
\begin{equation}
E_{\rm VIR-R,lim}(t)=5.0\,t^{-1/4}\;\rm W/cm^2.
\label{eq:iso_retina_E}
\end{equation}
For an exposure duration of $t=1800$ s, corresponding to 30 min, Eq.~\ref{eq:iso_retina_E} gives
\begin{equation}
E_{\rm VIR-R,lim}(1800~{\rm s}) \simeq 0.77\;\rm W/cm^2.
\end{equation}
For a diffuse retinal spot with an effective radius of 1 mm, the corresponding area is $A=\pi(0.1~\rm cm)^2\simeq3.14\times10^{-2}~\rm cm^2$, giving an optical-power value
\begin{equation}
P_{\rm retina,ISO}\simeq A E_{\rm VIR-R,lim}\simeq24~\rm mW.
\end{equation}
Applying a conservative factor-of-two design margin gives a retinal-power design value of approximately
\begin{equation}
P_{\rm retina,ISO}^{\rm design}\simeq12~\rm mW.
\end{equation}
This design margin is not used to redefine the standard limit; it is used only to account for speckle modulation, alignment uncertainty, calibration uncertainty, and patient-positioning variability.

For iris exposure over the relevant diameter $d_i$, the ISO 15004-2:2024 Group 1 visible/infrared thermal limit is related to the retinal limit for the corresponding exposure-duration regime. For the long continuous exposures considered here, the iris limit is more restrictive than the large-spot retinal limit. In the present conservative calculation, this gives a design value of approximately
\begin{equation}
P_{\rm iris,ISO}^{\rm design}\simeq6~\rm mW.
\end{equation}
This value is the most restrictive ISO 15004-2:2024 estimate for the diffuse Maxwellian configuration when the iris is exposed in the beam path.

For the cornea and crystalline lens, the relevant ISO 15004-2:2024 Group 1 limit for the visible/infrared anterior media is not the limiting condition for the present 852 nm diffuse Maxwellian configuration when compared with the iris estimate above. However, corneal and lenticular exposure should still be verified by aperture-averaged irradiance measurement at the corneal plane for the final instrument configuration.

\subsection{ANSI Z80.36-2021 assessment}
\label{sec:ansi2021_assessment}

ANSI Z80.36-2021 is the dedicated American standard for light-hazard protection of ophthalmic instruments. For a continuous-wave 852 nm source, the relevant Group 1 quantities are the weighted retinal visible/infrared thermal irradiance and, if applicable, the unweighted anterior-segment visible/infrared irradiance.

For retinal visible and infrared thermal hazard, the relevant weighted retinal irradiance limit is
\begin{equation}
E_{\rm VIR-R,lim}=0.7\;\rm W/cm^2.
\label{eq:ansi_retina}
\end{equation}
Using the same effective diffuse retinal area $A\simeq3.14\times10^{-2}~\rm cm^2$ gives
\begin{equation}
P_{\rm retina,ANSI}\simeq A E_{\rm VIR-R,lim}\simeq22~\rm mW.
\end{equation}
Applying a conservative factor-of-two design margin gives
\begin{equation}
P_{\rm retina,ANSI}^{\rm design}\simeq11~\rm mW.
\end{equation}
At this power, Eq.~\ref{eq:peak_pixel_scaling} gives a peak-pixel irradiance of approximately $126~\rm mW/cm^2$, before aperture averaging. The relevant aperture-averaged retinal irradiance is expected to be lower than this peak-pixel value for the diffuse profile.

ANSI Z80.36-2021 also defines an anterior-segment visible/infrared condition for instruments that, in normal use, irradiate the eye with a beam diameter of 2 mm or less at any point between the anterior corneal surface and the posterior crystalline-lens surface. The relevant unweighted anterior-segment irradiance is evaluated at the smallest beam diameter using a 1 mm-diameter averaging aperture. In the diffuse configuration, the measured beam waist is broader than the non-diffuse focus, and the anterior-segment condition is not the limiting condition in the same way as a narrow focused beam. Nevertheless, for conservative comparison with a 2 mm-diameter beam area, the irradiance associated with $P=11~\rm mW$ distributed over a 2 mm-diameter disk is
\begin{equation}
E_{\rm AS}(11~{\rm mW})\simeq \frac{11~\rm mW}{\pi(0.1~\rm cm)^2}\simeq0.35~\rm W/cm^2,
\end{equation}
well below the corresponding ANSI Z80.36-2021 anterior-segment visible/infrared Group 1 irradiance limit. Direct aperture-averaged measurement at the smallest anterior-segment beam diameter should be retained for final qualification.

\subsection{Recommended operating power for the present configuration}
\label{sec:recommended_power}

Table~\ref{tab:current_standards_summary} summarizes the current-standard assessment for the present diffuse Maxwellian illumination configuration. The limiting value is the ISO 15004-2:2024 iris estimate when the iris is included in the exposed path. We therefore recommend an operating power not exceeding
\begin{equation}
P_{\rm op}\leq6~\rm mW
\label{eq:recommended_power}
\end{equation}
at 852 nm, measured at the relevant illumination plane, for the present eyepiece, diffuser, and diffuse Maxwellian geometry. This value is intended as a conservative research-use operating level. It should be recalculated for any modification of wavelength, eyepiece focal length, diffuser angular profile, beam size, exposure duration, alignment procedure, or intended clinical use.

\begin{table*}[htbp]
\centering
\caption{Compact light-hazard assessment restricted to current ophthalmic standards for the diffuse Maxwellian illumination configuration at $\lambda=852$ nm. Values are design estimates based on measured diffuse beam profiles and conservative aperture assumptions. Final qualification should use the moving-aperture average required by each standard.}
\resizebox{\linewidth}{!}{%
\begin{tabular}{lllll}
\hline
Standard & Anatomical plane & Relevant hazard & Design estimate & Comment \\
\hline
ISO 15004-2:2024 & Retina & VIS/IR thermal, large spot & $\sim12$ mW & 30 min exposure, factor-of-two design margin \\
ISO 15004-2:2024 & Iris & VIS/IR thermal & $\sim6$ mW & Limiting ISO condition in the present geometry \\
ISO 15004-2:2024 & Cornea/lens & VIS/IR anterior media & Not limiting here & Verify by aperture-averaged measurement \\
ANSI Z80.36-2021 & Retina & VIS/IR thermal & $\sim11$ mW & Factor-of-two design margin \\
ANSI Z80.36-2021 & Anterior segment & VIS/IR irradiance & Not limiting here & Applies if beam diameter is $\leq2$ mm in anterior segment \\
\hline
\end{tabular}%
}
\label{tab:current_standards_summary}
\end{table*}

\section{Discussion}

In the absence of a diffuser (Fig.~\ref{fig_nodiffdistrib}), the illumination pattern near the beam focal plane contains a high-irradiance central hot spot. This localized irradiance concentration restricts the range of safe anterior-segment positioning and limits the use of Maxwellian-view illumination for wide-field retinal holography. Inserting an engineered diffuser spreads the input laser beam and filters the non-diffracted component, producing a broader and more homogeneous illumination distribution (Fig.~\ref{fig_diffdistrib}). As a result, the cornea--eyepiece distance can be reduced, increasing the reconstructed retinal field of view while avoiding the localized hot spot.

The Doppler images in Fig.~\ref{fig_DopplerImagesVsConfig} show that diffuse illumination preserves Doppler image rendering in both the broad fluctuation band and the high-frequency band. The diffuser does not remove the heterodyne detection principle: the backscattered retinal field still interferes with the reference beam, and temporal Doppler fluctuations remain accessible from high-speed interferogram sequences. The main optical change is the redistribution of illumination energy before ocular entry, rather than a change in the digital Doppler rendering method.

Restricting the light-hazard analysis to ISO 15004-2:2024 and ANSI Z80.36-2021 clarifies the manuscript and avoids parallel calculations from withdrawn or non-ophthalmic laser-safety frameworks. ISO 15004-2:2024 supersedes ISO 15004-2:2007 and reorganizes exposure criteria for Group 1 and Group 2 ophthalmic instruments. ANSI Z80.36-2021 is the relevant American ophthalmic light-hazard standard. For the present continuous-wave 852 nm source, both frameworks lead to a compact set of relevant checks: retinal visible/infrared thermal exposure, iris exposure when the iris is in the illumination path, and anterior-segment exposure when a small beam diameter occurs between the cornea and posterior crystalline lens.

The most conservative operating value obtained here is $P_{\rm op}\leq6~\rm mW$, set by the ISO 15004-2:2024 iris estimate for the present diffuse Maxwellian geometry. This value is higher than the power typically used in earlier focused configurations while retaining a safety margin based on measured diffuse beam profiles. It is also compatible with the qualitative Doppler image results shown here. The exact limiting condition can change with the eyepiece focal length, diffuser angular distribution, patient alignment, pupil diameter, exposure duration, and whether the iris is intentionally or incidentally illuminated.

The proposed configuration can therefore be used effectively for Maxwellian-view retinal Doppler holography. In the tested configuration, posterior-pole imaging was achieved without pharmacological dilation. More generally, diffuse Maxwellian illumination may reduce the need for dilation, depending on pupil diameter, fixation, field-of-view requirements, and clinical protocol.

Diffuse illumination is also relevant beyond wide-field Maxwellian imaging. In small-field or high-resolution retinal laser Doppler imaging, local retinal irradiance can become restrictive if a collimated or tightly focused coherent beam forms a small retinal image. By reducing spatial coherence and redistributing energy, diffuse illumination can mitigate local hot spots, similarly in spirit to approaches using multimode fibers to control spatial coherence in coherent ophthalmic imaging \cite{Auksorius2021}. A dedicated study will be required to assess the optimal balance between spatial resolution, coherence, Doppler sensitivity, and light-hazard limits in high-resolution implementations.

The present study has limitations. First, the light-hazard assessment is a design calculation based on measured beam profiles and conservative geometrical assumptions; it is not a regulatory certification. Second, the analysis should be repeated with a moving-aperture averaging algorithm applied directly to calibrated irradiance maps at each relevant anatomical plane. Third, Doppler image preservation is demonstrated qualitatively; future work should quantify the field of view, vessel-to-background contrast, Doppler signal-to-noise ratio, and usable retinal area across subjects and pupil sizes.

\section{Conclusion}

We demonstrated that engineered diffuse illumination enables Maxwellian-view retinal Doppler holography while avoiding the localized laser hot spot produced by focused non-diffuse illumination. In the tested configuration, the diffuser allowed the cornea--eyepiece distance to be reduced, thereby increasing the digitally reconstructed retinal field of view without degrading Doppler image rendering in the analyzed fluctuation bands.

The light-hazard analysis was restricted to the current ophthalmic standards ISO 15004-2:2024 and ANSI Z80.36-2021. For the present 852 nm diffuse Maxwellian configuration, the most conservative design estimate is an operating power not exceeding approximately 6 mW, measured at the relevant illumination plane, with final qualification to be performed by aperture-averaged irradiance measurements for the final instrument configuration. These results support diffuse Maxwellian illumination as a practical route toward safer, wider-field, non-mydriatic Doppler holography of the human retina.

The authors declare that the research was conducted in the absence of any commercial or financial relationship that could be construed as a potential conflict of interest.

\section{Acknowledgements}

This work was supported by the IHU FOReSIGHT (ANR-18-IAHU-01), the European Research Council, the Sesame program of the Region Ile-de-France (4DEye, ANR-10-LABX-65), and the French National Research Agency (ANR LIDARO).

\bibliographystyle{unsrt}

\begin{thebibliography}{10}

\bibitem{Puyo2018}
L. Puyo, M. Paques, M. Fink, J.-A. Sahel, and M. Atlan.
\newblock In vivo laser Doppler holography of the human retina.
\newblock \emph{Biomedical Optics Express}, 9(9):4113--4129, Sep. 2018.

\bibitem{Puyo2021}
L. Puyo, C. David, R. Saad, S. Saad, J. Gautier, J.-A. Sahel, V. Borderie, M. Paques, and M. Atlan.
\newblock Laser Doppler holography of the anterior segment for blood flow imaging, eye tracking, and transparency assessment.
\newblock \emph{Biomedical Optics Express}, 12(7):4478--4495, 2021.

\bibitem{Puyo2020}
L. Puyo, M. Paques, and M. Atlan.
\newblock Spatio-temporal filtering in laser Doppler holography for retinal blood flow imaging.
\newblock \emph{Biomedical Optics Express}, 11:3274--3287, 2020.

\bibitem{Sliney2005}
D. Sliney et al.
\newblock Adjustment of guidelines for exposure of the eye to optical radiation from ocular instruments: statement from a task group of the International Commission on Non-Ionizing Radiation Protection (ICNIRP).
\newblock \emph{Applied Optics}, 44(11):2162--2176, 2005.

\bibitem{PCA2020}
L. Puyo, L. Bellonnet-Mottet, A. Martin, F. Te, M. Paques, and M. Atlan.
\newblock Real-time digital holography of the retina by principal component analysis.
\newblock arXiv:2004.00923, 2020.

\bibitem{Spector1990}
R. H. Spector.
\newblock Clinical Methods: The History, Physical, and Laboratory Examinations, 3rd edition, Chapter 58: The pupils.
\newblock National Center for Biotechnology Information, 1990.

\bibitem{ISO2024}
International Organization for Standardization.
\newblock ISO 15004-2:2024, Ophthalmic instruments -- Fundamental requirements and test methods -- Part 2: Light hazard protection.
\newblock 2024.

\bibitem{ANSI_Z80_2021}
American National Standards Institute.
\newblock ANSI Z80.36-2021, American National Standard for Ophthalmics -- Light Hazard Protection for Ophthalmic Instruments.
\newblock 2021; revised February 9, 2022.

\bibitem{Auksorius2021}
E. Auksorius, D. Borycki, and M. Wojtkowski.
\newblock Multimode fiber enables control of spatial coherence in Fourier-domain full-field optical coherence tomography for in vivo corneal imaging.
\newblock \emph{Optics Letters}, 46:1413--1416, 2021.

\end{thebibliography}

\end{document}